\documentclass{article}
\usepackage{bm,latexsym,amsmath,amssymb,amsfonts,fancyhdr,color,graphicx,multirow,slashed,cite,bbold}
\usepackage[a4paper,bottom=3cm,top=2.5cm,head=0mm,width=17cm,dvipdfm]{geometry}
\usepackage[usenames,dvipsnames,svgnames,table]{xcolor}
\usepackage[colorlinks=true,
            linkcolor=blue,
            urlcolor=blue,
            citecolor=green,          
						bookmarks=true,
						bookmarksnumbered=true,
						breaklinks=true,
						pdfstartview=FitBH]{hyperref}
\usepackage[dotinlabels]{titletoc}
\usepackage{titlesec}
\usepackage{authblk,ulem}
\usepackage{multirow}
\usepackage{makecell}
\usepackage{cancel}
\usepackage{enumitem}
\numberwithin{equation}{section}
\allowdisplaybreaks[4]

\titlelabel{\thetitle.\quad \hspace{-0.8em}}
\titlecontents{section}
              [1.5em]
              {\vspace{4mm} \large \bf}
              {\contentslabel{1em}}
              {\hspace*{-1em}}
              {\titlerule*[.5pc]{.}\contentspage}
\titlecontents{subsection}
              [3.5em]
              {\vspace{2mm}}
              {\contentslabel{1.8em}}
              {\hspace*{.3em}}
              {\titlerule*[.5pc]{.}\contentspage}
\titlecontents{subsubsection}
              [5.5em]
              {\vspace{2mm}}
              {\contentslabel{2.5em}}
              {\hspace*{.3em}}
              {\titlerule*[.5pc]{.}\contentspage}

\newcommand{\titledef}{Neutrino CP Measurement in the Presence of RG Running with Mismatched Momentum Transfers} % Insert Title here!!!
\hypersetup{ pdfauthor = {Shao-Feng Ge},
	     pdftitle = {\titledef}, % Insert title here!!!
	     pdfsubject = {}, % Insert subject here!!!
             pdfkeywords = {}, % Insert keywords here!!!
	     pdfcreator = {LaTeX with hyperref package},
	     pdfproducer = {dvips + ps2pdf} }

\usepackage{stackengine}

\definecolor{gesfblack}{rgb}{0,0,0}

\definecolor{gesfblue}{rgb}{0.08,0.42,0.76}

\definecolor{gesfgreen}{rgb}{0,1,0}

\definecolor{gesfgrey}{rgb}{0.5,0.5,0.5}

\definecolor{gesflanse}{rgb}{0.00,0.50,0.50}

\definecolor{gesfpurple}{rgb}{0.47,0.19,0.42}

\definecolor{gesfred}{rgb}{1,0,0}

\definecolor{gesfwhite}{rgb}{1,1,1}

\definecolor{gesfyellow}{rgb}{0.7,0.4,0.3}

\newcommand{\gsec}[1]{{\hypersetup{linkcolor=red}Sec.\,\ref{#1}\hypersetup{linkcolor=blue}}}

\newcommand{\geqn}[1]{\hypersetup{linkcolor=blue}Eq.\,(\ref{#1})\hypersetup{linkcolor=blue}}
\newcommand{\gfig}[1]{{\hypersetup{linkcolor=violet}Fig.\,\ref{#1}\hypersetup{linkcolor=blue}}}
\newcommand{\gtab}[1]{{\hypersetup{linkcolor=gesflanse}Table~\ref{#1}\hypersetup{linkcolor=blue}}}

\definecolor{Orange}{cmyk}{0,0.61,0.87,0}
\definecolor{JungleGreen}{cmyk}{0.99,0,0.52,0}
\definecolor{OliveGreen}{cmyk}{0.64,0,0.95,0.40}
\definecolor{Brown}{cmyk}{0,0.81,1,0.60}
\definecolor{RoyalBlue}{cmyk}{0.71,0.53,0,0.12}
\definecolor{Gray}{cmyk}{0,0,0,0.40}
\definecolor{LightPink}{cmyk}{0.0,0.25,0,0}
\definecolor{LLightPink}{cmyk}{0.0,0.10,0,0}
\definecolor{LightBlue}{cmyk}{0.25,0,0,0}
\definecolor{LightGray}{cmyk}{0,0,0,0.2}

\usepackage{float}

\graphicspath{{FinalFigures/}}

\begin{document}
\fontsize{12pt}{14pt}\selectfont

\title{%\begin{flushright}
       %\mbox{\normalsize IPMU18-xxxx}
       %\end{flushright}
			 %\vskip 20pt
       \textbf{\Large \titledef}} % Insert title here!!!
\author[1,2]{{\large Shao-Feng Ge} \footnote{\href{mailto:gesf@sjtu.edu.cn}{gesf@sjtu.edu.cn}}}
\affil[1]{Tsung-Dao Lee Institute \& School of Physics and Astronomy, Shanghai Jiao Tong University, Shanghai 200240, China}
\affil[2]{Key Laboratory for Particle Astrophysics and Cosmology (MOE) \& Shanghai Key Laboratory for Particle Physics and Cosmology, Shanghai Jiao Tong University, Shanghai 200240, China}
\author[1,2]{{\large Chui-Fan Kong} \footnote{\href{mailto:kongcf@sjtu.edu.cn}{kongcf@sjtu.edu.cn}}}
\author[3]{{\large Pedro Pasquini} \footnote{\href{mailto:pedrosimpas@g.ecc.u-tokyo.ac.jp}{pedrosimpas@g.ecc.u-tokyo.ac.jp}}}
\affil[3]{Department of Physics, University of Tokyo, Bunkyo-ku, Tokyo 113-0033, Japan}
\date{\today}

\maketitle

\begin{abstract}
\fontsize{12pt}{14pt}\selectfont

The neutrino mixing parameters are 
expected to have renormalization group (RG) running effect in the
presence of new physics.
If the momentum transfers at production and
detection mismatch with each other, the oscillation
probabilities are generally modified and become dependent
on not just the neutrino energy but also the momentum transfer.
Even in the limit of vanishing baseline, the
transition probability for the appearance channel
is interestingly not zero. This would significantly
affect the sensitivity of the genuine leptonic Dirac
CP phase. We further explore
the possibility of combining the long- and short-baseline 
neutrino experiments to constrain such RG
running effect for the purpose of guaranteeing
the CP measurement. To simulate the double dependence
on the neutrino energy and momentum transfer,
we extend the usual GLoBES simulation of
fixed baseline experiments and use a two-dimensional 
$\chi^2$ analysis to obtain sensitivities.
\end{abstract}

%\tableofcontents

\newpage

\section{Introduction}

\label{sec:intro}
The matter-antimatter asymmetry in the Universe 
is one of the most fundamental questions \cite{Canetti:2012zc}.
To explain this asymmetry, the charge-parity (CP)
violation is necessary \cite{Sakharov:1967dj}.
In the Standard Model (SM) of particle physics, the quark 
mixing matrix introduces a CP phase but the induced baryon
asymmetry is too small to account for the observed value 
\cite{Gavela:1993ts,Huet:1994jb,Gavela:1994dt}.
In the extended SM with heavy right-handed neutrinos, 
it has been shown that the leptonic CP violation
could generate sufficient matter–antimatter disparity
through the leptogenesis mechanism
\cite{Fukugita:1986hr,Buchmuller:2005eh,Davidson:2008bu}.
It is interesting to observe that the leptonic CP
phases in the neutrino mixing matrix \cite{PDG22-NuCP}
provide potential sources of CP violation that can
be directly used to explain the matter-antimatter asymmetry 
\cite{Pascoli:2006ie,Branco:2006ce,Pascoli:2006ci,
Anisimov:2007mw,Molinaro:2009lud,Ge:2010js,Dolan:2018qpy}.

The leptonic Dirac CP phase $\delta_D$ of the Pontecorvo-Maki-Nakagawa-Sakata (PMNS)
matrix can manifest itself in neutrino oscillation
and hence can be measured therein. Its first measurement
results were published in 2019
\cite{NOvA:2019cyt, T2K:2019bcf}
by the two 
long-baseline (LBL) neutrino oscillation 
experiments NO$\nu$A \cite{NOvA:2004blv} and T2K \cite{T2K:2011qtm}.
With the normal ordering (NO) of neutrino masses
that is preferred by the latest results, the 2021 NO$\nu$A
result gives $\delta_D=148^\circ{}^{+49^\circ}_{-157^\circ}$  
\cite{NOvA:2021nfi} while the 2023 T2K 
result has
$\delta_D = 247^\circ{}^{+56^\circ}_{-36^\circ}$ 
\cite{T2K:2023smv}.
The next-generation LBL experiments T2HK \cite{Abe:2011ts}
and DUNE \cite{DUNE:2015lol} under construction are expected
to greatly improve the precision on $\delta_D$.
With about 10 years of data taking, each  
experiment can reach 
$\sim 10^\circ$ precision level \cite{Hyper-Kamiokande:2018ofw,DUNE:2021mtg}. 
In addition, ESS$\nu$SB \cite{ESSnuSB:2021azq}
is sensitive to $\delta_D$ by detecting  
neutrinos at the second oscillation peak, 
while T2HKK \cite{Hyper-Kamiokande:2016srs} 
and MOMENT \cite{Tang:2019wsv} are sensitive 
to both the first and second oscillation 
maxima. With around twice the energy and 
baseline of DUNE, the P2O experiment \cite{Akindinov:2019flp}
is sensitive to both $\delta_D$ and the
matter effect. Additional low-energy neutrinos 
from muon decay at rest ($\mu$DAR) provide
complementary measurements to accelerator LBL experiments,
such as DAE$\delta$ALUS \cite{Alonso:2010fs},
TNT2HK \cite{Evslin:2015pya},
$\mu$DARTS \cite{Ciuffoli:2014ika,Ciuffoli:2015uta},
DAE$\delta$ALUS+JUNO \cite{Smirnov:2018ywm},
and DUNE+$\mu$THEIA \cite{Ge:2022iac}.
Besides accelerator neutrinos,
CP measurement with sub-GeV atmospheric 
neutrino oscillation is also proposed for 
Super-PINGU \cite{Razzaque:2014vba,Razzaque:2015fea},
Super-ORCA \cite{Hofestadt:2019whx}, and even 
at JUNO \cite{JUNO:2015zny} or
DUNE \cite{Kelly:2019itm}.

Although the neutrino oscillation measurement has
entered the precision era 
\cite{deSalas:2020pgw,Gonzalez-Garcia:2021dve}, 
the interpretation of the experimental data to extract
the leptonic Dirac CP phase $\delta_D$ still faces
various theoretical challenges.
On the SM side, the matter density uncertainty 
can greatly reduce the sensitivity to $\delta_D$, 
especially when the value of $\delta_D$ is around
the maximal values $90^\circ$ or $270^\circ$
\cite{DeRomeri:2016qwo,Raut:2017dbh, 
Kelly:2018kmb, King:2020ydu,Ge:2022iac}.
Beyond the SM, there are various new physics scenarios
that are capable of mimicking the CP violation 
effect and hence disturbing the CP 
measurement. For example, the neutrino 
non-standard interaction (NSI) 
contributes extra matter potential
\cite{Farzan:2017xzy,Proceedings:2019qno} 
that receives different sign between 
neutrino and anti-neutrino modes to fake CP
and significantly reduce the CP phase sensitivity 
\cite{Ge:2016dlx,GeProceedings,Bakhti:2020fde,Chatterjee:2021wac,
Medhi:2021wxj,Medhi:2022qmu}. In addition, the 
Lorentz invariance violation (LIV) 
\cite{Diaz:2014yva,Torri:2020dec} 
induces an effective Hamiltonian with 
both CPT-odd and CPT-even terms that have similar
form as the effective Hamiltonian induced by NSI.
Consequently, the effective Hamiltonian will
modify the neutrino oscillation probability and hence
impact on the CP phase sensitivity just like NSI
\cite{Majhi:2019tfi,KumarAgarwalla:2019gdj,Lin:2021cst,
Fiza:2022xfw}. Moreover, the non-unitary mixing
due to heavy neutrinos \cite{Forero:2011pc,Escrihuela:2015wra} 
allows extra CP phases to fake the genuine CP effect
\cite{Ge:2016xya,GeProceedings,Pasquini:2017zwk,Escrihuela:2016ube,Forero:2021azc,Martinez-Soler:2018lcy}.

Besides the new physics scenarios listed above, 
both the SM and
the BSM interactions can generate less explored
effects in neutrino oscillation such as  
the renormalization group (RG) running of 
the neutrino mixing parameters
\cite{Huang:2023nqf, Casas:1999tg,Chankowski:2001mx,Antusch:2003kp,Ray:2010rz,Ohlsson:2013xva}. 
Various neutrino mass models, such as the canonical 
seesaw models, the inverse seesaw model, the 
scotogenic model, and the radiative Dirac model 
can lead to RG running effects
\cite{Ray:2010rz,Ohlsson:2013xva}.
Consequently, the neutrino mass and mixing parameters
evolve as functions of the relevant energy scale.
With mismatched energy scales between the neutrino
production and detection processes, the neutrino
RG running can affect its oscillation probabilities
\cite{Babu:2021cxe}.

In our paper, we carefully study how RG
running can affect the leptonic CP phase measurement. In 
\gsec{sec:EDependentOsci}, we review the neutrino
RG running features and introduce a model-independent
parametrization for the running behaviors. We then 
establish in \gsec{sec:Q2dependentP} the general
formalism of neutrino oscillations in the presence
of RG running. In particular, we adopt the
quantum field theory (QFT) language and compare
with the usually adopted quantum mechanics (QM) approach.
With higher energy, DUNE
can provide larger variation at the relevant energy
scale than other experiments to allow more significant
RG effects. In \gsec{sec:4}, we  
elaborate 
how a two-dimensional neutrino oscillation simulation,
including neutrino energy and momentum transfer
reconstructions, should be established for GLoBES
with DUNE as
a specific realization. The interplay between RG
running and the genuine Dirac CP phase is evaluated
with the extended two-dimensional $\chi^2$ analysis
to give the projected sensitivities at DUNE. Our conclusion and discussion about this
new effect can be found in 
\gsec{sec:conclusion}.

\section{RG Running of Neutrino Mixing Parameters}
\label{sec:EDependentOsci}

Although neutrino oscillation is the first
new physics supported by various experimental
measurements, we are still not so sure what is
really happening behind the scene. The new physics
that we have been looking for may not be as
appearant as the non-standard interactions (NSI)
that appears at tree level already and
affects the neutrino oscillation behavior, but
make itself manifest via loop correction and the RG
running effect. Studying the RG running effect
in neutrino oscillation is actually an inevitable
way of exploring new physics.

The neutrino mixing matrix $U$ is defined according
to the neutrino mass matrix diagonalization,
$U^\dagger M^\dagger_\nu M_\nu U={\rm Diag}(m^2_1,m^2_2,m^2_3)$.
In other words, the RG running effect of the neutrino 
mixing is inherited from the running of the 
neutrino mass matrix. On the model building side,
the neutrino mass is typically
generated at a high scale while the mass  
value and the corresponding neutrino mixing 
parameters are measured experimentally at a low 
scale. Across different energy scales, the RG
evolution effects need to be included.

The RG running of neutrino mixing
can be generally parameterized using the
$\beta$ function,
\begin{equation}
  \frac{d \mathcal O_\nu}{d\ln |Q^2|}
\equiv
  \beta_\nu,
\label{eq:evolution_eq_Unu}
\end{equation}
where $|Q^2|$ is the absolute value of momentum transfer
\cite{Babu:2021cxe}. Choosing the Lorentz-invariant
momentum transfer $|Q^2|$ as the RG running variable is
widely used in the literature and known as the Gell Mann-Low
scheme \cite{Gell-Mann:1954yli, Wu:2013ei}. Such a choice is
consistent with the previous studies about the RG running effect
in neutrino oscillations
\cite{Bustamante:2010bf, Babu:2021cxe,Babu:2022non}.
For generality, we use $\mathcal O_\nu$ to denote the 
neutrino observables such as the mixing angles and the 
Dirac CP phase. In neutrino oscillation, the 
neutrino propagates mainly as the on-shell mass eigenstates and
the RG scale of the propagator is fixed as neutrino masses,
$Q^2=m^2_{\nu}$. Consequently, the RG running contribution 
from the neutrino
propagator part can be ignored \cite{Babu:2021cxe}.  
In other words, the RG running of the neutrino mixing matrix
should come from the charged-current vertex part
\cite{Bustamante:2010bf, Babu:2021cxe,Babu:2022non}. 
Both the neutrino production and the detection 
processes are affected by RG running via modification of 
the mixing angles and coupling constants with different
scales. Below the new physics scale, the RG evolution
determined by the SM
\cite{Chankowski:1993tx,Babu:1993qv} is negligibly
small \cite{Ohlsson:2013xva} and hence the
mixing parameters are treated as constants.
Any sizable RG running effect should appear
above the new physics scale.
The neutrino masses can also 
run in a similar way as \geqn{eq:evolution_eq_Unu}. 
However, the impact of running neutrino masses is 
negligible \cite{Casas:1999tg,Chankowski:2001mx,Antusch:2003kp,Ray:2010rz,Ohlsson:2013xva,Babu:2021cxe}
for the oscillation process of interest 
and hence omitted in this paper for simplicity.

For generality, we simply treat $\beta_\nu$ as a 
perturbative constant. In this sense, the solution of
\geqn{eq:evolution_eq_Unu} for the
three mixing angles ($\theta_s \equiv \theta_{12}$,
$\theta_a \equiv \theta_{23}$, and
$\theta_r \equiv \theta_{13}$)
and the Dirac CP phase $\delta_D$ can be 
generally expressed as,
\begin{align}
  \theta_{ij}(Q^2)
\equiv 
  \theta_{ij}(Q^2_0)
+ \beta_{ij}
  \ln\left(\left|\frac{Q^2}{Q^2_0}\right|\right),
\quad \mbox{and} \quad
  \delta_D(Q^2)
\equiv 
  \delta_D(Q^2_0)
+ \beta_\delta 
  \ln\left(\left|\frac{Q^2}{Q^2_0}\right|\right).
\label{eq:new_parametrization}
\end{align}
The reference value $|Q^2_0|$ corresponds to the
new physics scale and the evolution formulas
apply for $|Q^2| > |Q^2_0|$ while reducing to
constant values for $|Q^2| < |Q^2_0|$. 
If the new physics scale is much higher 
than the experimental scales, the running 
effect will not be measured at the current 
low energy neutrino experiments. 
To test the running effect
at the $\mathcal{O}({\rm GeV})$ LBL 
experiments such as DUNE, we take $Q^2_0 = 1$\,MeV 
as a benchmark value throughout this paper.
Furthermore, we need four parameters to describe 
the evolution, three $\beta_{ij}$ for the mixing 
angles and one $\beta_\delta$ for the Dirac CP 
phase. The RG running effect can lead to 
non-standard oscillation behaviors as we 
elaborate below.

\section{General Formalism of Neutrino Oscillation}
\label{sec:Q2dependentP}

\subsection{QFT Description of Neutrino Oscillation}

\begin{figure}[t]
\centering
\includegraphics[width=0.98\textwidth]{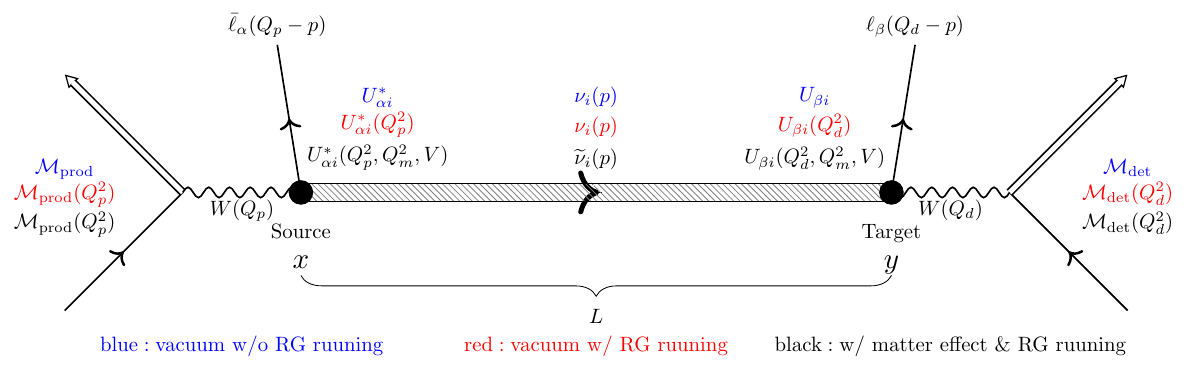}
\caption{The schematic plot for the QFT description of neutrino 
oscillation including three cases:
1) vacuum oscillation
without considering RG running (blue); 
2) vacuum oscillation with RG running (red);
as well as
3) oscillation with both RG running and  
matter effect (black). 
}
\label{fig:QFT_nuOSc}
\end{figure}

We start with the vacuum neutrino oscillation
without considering RG running which is conceptually
clearer to set the stage in the QFT framework.
A typical neutrino oscillation experiment 
involves the neutrino production, propagation, and
detection processes. For the case of neutrino (rather than
anti-neutrino) oscillation, a charged anti-lepton $\bar 
\ell_\alpha$ appears at the position
$x \equiv (x_0,{\boldsymbol x})$
when producing an $\alpha$-flavor neutrino $\nu_\alpha$
while a charged lepton $\ell_\beta$ is observed at the position
$y \equiv (y_0,{\boldsymbol y})$ for the detection of a $\beta$-flavor neutrino $\nu_\beta$ as shown in
\gfig{fig:QFT_nuOSc}.
In between, neutrinos propagate as mass eigen-fields
($\nu_i \equiv \sum_\alpha U^\ast_{\alpha i}\nu_\alpha$)
over a macroscopic spatial distance
${\boldsymbol L}\equiv {\boldsymbol y} - {\boldsymbol x}$.
The full transition matrix element 
$\mathcal{T}$ for the above 
processes can be read off directly from 
\gfig{fig:QFT_nuOSc} \cite{Grimus:1996av},
\begin{eqnarray}
  \mathcal T
\equiv
  \sum_i
  \int d^4x d^4y
  \mathcal M^\mu_{\rm det}(y)
  \frac {i g}{\sqrt 2} U_{\beta i}
  \bar u_\beta \gamma_\mu
  P_L
  \int \frac{d^4 p}{(2\pi)^4}
  \frac{i(\slashed p+m_i)e^{i p \cdot (x-y)}}
       {p^2-m^2_i+i\epsilon}
  \frac {i g}{\sqrt 2} 
  U^*_{\alpha i}\gamma_\nu P_L v_\alpha
  \mathcal M^\nu_{\rm prod}(x),
\quad
\label{eq:M}
\end{eqnarray}
where the production and detection matrix 
elements $\mathcal M^\mu_{\rm prod}(x)$ 
and $\mathcal M^\nu_{\rm det}(y)$ 
contain those relevant interactions except
the charged-current vertex between neutrino
and charged lepton. To also take the associated
charged leptons into consideration,
the charged current (CC) interaction mediated by
the SM $W^\pm$ boson with weak gauge coupling
$g$ should be involved.
With $\overline \psi$ to produce a neutrino,
the associated matrix element $U^*_{\alpha i}$
receives a complex conjugation. Similarly,
the detection part annihilates the intermediate
neutrino with $\psi$ to have $U_{\beta i}$.
The projector $P_L$ selects the left-handed
fermions.

The oscillation behavior as a periodic function
of the baseline $L$ is encoded in the propagation
phase factor $e^{i p \cdot (x - y)}$ from $x$ to $y$.
For the mass eigenstate $\nu_i$ propagator,
\begin{align}
  \int \frac{d^4 p}{(2\pi)^4}
  \frac{i(\slashed p+m_i)e^{i p\cdot (x-y)}}
       {p^2-m^2_i+i\epsilon}
=
  \int \frac{d^4 p}{(2\pi)^4}
  \frac{i\sum_{s} u_i(p,s)\bar u_i(p,s)e^{i p\cdot (x-y)}}
       {p^2-m^2_i+i\epsilon},
\label{eq:propagator}
\end{align}
its numerator can be replaced by two neutrino spinors
according to the spin sum relation, 
$\slashed p+m_i=\sum_{s} u_i(p,s)\bar u_i(p,s)$.
In the ultra-relativistic limit, only the
left-handed chiral component can contribute
since the right-handed counterpart is projected out
by
$P_L u(p,+)=0$. Then the neutrino spinors
$u_i(p,-) \bar u_i(p,-)$ in the numerator 
complete the matrix elements
for production and detection,
\begin{subequations}
\begin{align}
  \mathcal M_{\rm prod}(x)
& \equiv
  \frac {i g}{\sqrt 2} \bar u_i(p,-) \gamma_\nu P_L v_\alpha
  \mathcal M^\nu_{\rm prod}(x),
\\
  \mathcal M_{\rm det}(y)
& \equiv
  \frac {i g}{\sqrt 2}
  \mathcal M^\mu_{\rm det}(y)
  \bar u_\beta \gamma_\mu P_L u_i(p,-).
\label{eq:Mdp}
\end{align}
\end{subequations}
Since the neutrino mass is typically much smaller
than the momentum transfer in production and 
detection, their matrix elements
$\mathcal M_{\rm prod}$ and 
$\mathcal M_{\rm det}$ are essentially independent
of the neutrino masses. Consequently, the sum over 
the mass eigenstantes involves only 
the neutrino propagator, 
\begin{align}
  \mathcal{T}
=
  \int \frac{d^4p}{(2\pi)^4}
  d^4x d^4y 
  \mathcal M_{\rm det}(y)
\left[
\sum_i
  U_{\beta i}
  \frac{ie^{ip\cdot (x-y)}}
       {p^2-m^2_i+i\epsilon}
  U^\ast_{\alpha i}
\right]
  \mathcal M_{\rm prod}(x).
\end{align}

The integral over the four-momentum 
$p$ is divided into two parts, the 
integration over the three momentum
$\boldsymbol p$ and energy $p_0$,
respectively. The three-momentum 
integration can be calculated
using the Grimus-Stockinger theorem \cite{Grimus:1996av},
\begin{eqnarray}
  \int \frac{d^3 {\boldsymbol p}}{(2\pi)^3} 
  f({\boldsymbol p}) \frac{e^{i {\boldsymbol p}\cdot {\boldsymbol L}}}{p^2_0 - {\boldsymbol p}^2 -m^2_i+i\epsilon }
\xrightarrow{|{\boldsymbol L}|\rightarrow \infty}  
  \left. 
  - \frac{1}{4\pi L} 
    f\left(|{\boldsymbol p}|\hat{\boldsymbol L}\right)
    e^{i |{\boldsymbol p}| L}
    \right|_{|{\boldsymbol p}| = \sqrt{p_0^2 - m_i^2}},
\label{eq:GStheorem}
\end{eqnarray}
with the exponential term being $e^{-i{\boldsymbol p}\cdot ({\boldsymbol x}-{\boldsymbol y})}=e^{i{\boldsymbol p}\cdot {\boldsymbol L}}$.
The above equation means that the leading 
contribution is given by neutrinos on 
mass shell, $|{\boldsymbol p}|
\approx p_0 - m_i^2 / 2 p_0$, and propagating
from source to detector in the direction of
$\hat{\boldsymbol L} = 
{\boldsymbol L}/|{\boldsymbol L}|$. Most
importantly, the full transition matrix element,
\begin{eqnarray}
  \mathcal{T}
\propto
    \frac{1}{L}
    \int
    d p_0
    d^4x d^4y 
  \mathcal M_{\rm det}(y)
\left(
  \sum_i U_{\beta i} e^{-i\frac{m_i^2}{2p_0}L} U^*_{\alpha i}
\right)
  \mathcal M_{\rm prod}(x),
\label{eq:M-final}
\end{eqnarray}
is inversely proportional to the baseline $L$
which becomes a flux dilution factor $1/L^2$
in $|\mathcal T|^2$ due to neutrino number conservation in 
the three-dimensional space.
The two spatial and one energy $p_0$ integrations
result in a $\delta-$function for energy
conservation of the whole process \cite{Grimus:1996av}
with explicit calculation and more details to be
found in \cite{Giunti:1993se,Grimus:1996av,
Beuthe:2001rc,Falkowski:2019kfn}. 
For macroscopic distance, the total 
process decomposes into the neutrino flux from
production, the neutrino scattering cross section
of detection, and the neutrino
oscillation probability.

The neutrino oscillation probability,
$P_{\nu_\alpha \rightarrow \nu_\beta}$,
is then defined
as the absolute value squared of the
amplitude in
\geqn{eq:M-final},
\begin{align}
  P_{\alpha\beta}
\equiv
  \left| 
  \mathcal A_{\beta\alpha}^{\rm osc}
  \right|^2,
\quad {\rm with} \quad 
  \mathcal A_{\beta\alpha}^{\rm osc}
\equiv 
  \sum_i 
  U_{\beta i}
  e^{-i\frac{m_i^2}{2E_\nu}L}
  U^*_{\alpha i},
  \label{eq:QFT-A}
\end{align}
by extracting those irrelevant terms for the oscillation
probability to reduce to a $\delta$ function,
$P_{\alpha \beta} (L = 0) = \delta_{\alpha \beta}$.
We can see that the amplitude
$\mathcal A^{\rm osc}_{\beta \alpha}$
is exactly the parenthesis of \geqn{eq:M-final}.

Note that the above QFT formalism for neutrino
oscillation has two major differences with its quantum
mechanics (QM) counterpart. In QM, the $\alpha$ flavor
neutrino 
$\nu_\alpha$ is described by a quantum state 
$|\nu_\alpha\rangle \equiv a^\dagger_\alpha|0\rangle$
and similarly for the mass eigenstate
$|\nu_i\rangle\equiv a_i^\dagger|0\rangle$.
Since $\nu_\alpha\equiv \sum_i U_{\alpha i}\nu_i$, 
the neutrino operators $a_\alpha$ and $a_i$
(not $b_\alpha$ or $b_i$) have the same similarity
transformation $a_\alpha =\sum_i U_{\alpha i} a_i$
according to $\nu \sim a + b^\dagger$. Then, the
similarity transformation on the quantum states,
$|\nu_\alpha \rangle = \sum_i U^*_{\alpha i} |\nu_i\rangle$,
receives an extra complex conjugation.
The second difference appears in
the oscillation phase factor, $e^{-i E_i t}$,
which has a minus sign according to the Schrodinger
equation while its counterpart from the neutrino
propagator in QFT is $e^{i|\mathbf{p}|L}$ without
minus sign. With simultaneous presence of both
differences, the oscillation probability derived
in both QFT and QM formalism is the same.

\subsection{Vacuum Oscillation with RG Running and Zero-Distance Effect}
\label{sec:vac-zd}

Note that physical variables such as the 
neutrino mixing parameters and masses depend on
momentum transfer in the presence of RG running 
as mentioned in \gsec{sec:EDependentOsci}. 
Especially for a typical neutrino oscillation 
experiment, the momentum transfers in neutrino
production, propagation, and detection processes
are different from each other. 
However, since the neutrino mass receives 
negligible RG running effect
\cite{Casas:1999tg,Chankowski:2001mx,Antusch:2003kp,Ray:2010rz,Ohlsson:2013xva,Babu:2021cxe},
the neutrino mass eigen-fields can be treated
universally during the whole process.
In this sense, the neutrino flavor eigen-fields 
become momentum transfer dependent via the mixing
matrix $U(Q^2)$, $\nu_\alpha(Q^2) 
\equiv \sum_i U_{\alpha i}(Q^2)\nu_i$. 
Besides the neutrino mixing matrices and 
neutrino flavor eigen-fields, 
quantum corrections can also
generate a RG running effect 
on the gauge coupling $g$
of CC weak interactions, $g\rightarrow g(Q^2)$
\cite{Babu:2021cxe}. 
 As shown 
in \gfig{fig:QFT_nuOSc}, the full transition
matrix element for the vacuum oscillation with
RG running effect can be obtained by 
replacing the relevant parameters in 
\geqn{eq:M-final} with their $Q^2$-dependent 
counterparts, 
\begin{align}
  \mathcal T
=
  -\frac{i}{8\pi^2 L}
    \int
    d p_0
    d^4x d^4y 
  \mathcal M_{\rm det}(y,Q^2_d)
  \left[
  \sum_i 
  U_{\beta i}(Q^2_d)
  e^{-i\frac{m_i^2}{2E_\nu}L}
  U^*_{\alpha i}(Q^2_p)
  \right]
  \mathcal M_{\rm prod}(x,Q^2_p),
\label{eq:M-Q2}
\end{align}
where the momentum transfers in the 
neutrino production and detection processes
are defined as $Q^2_p$ and $Q^2_d$.
Finally, the vacuum oscillation amplitude 
with the RG running effect
receives a modification on the mixing 
matrix from
\geqn{eq:QFT-A},
\begin{eqnarray}
  \mathcal{A}^{\rm osc}_{\beta \alpha}
= 
  U_{\beta i} (Q^2_d)
  e^{- i L \frac{m^2_i}{2E_\nu}}
  U^\ast_{\alpha i}(Q^2_p).
\label{eq:amplitudeQFT}
\end{eqnarray}
The oscillation amplitude and consequently
the corresponding probability $P_{\alpha \beta}(E_\nu, Q^2_p, Q^2_d)$ have 
dependence on the production and detection
momentum transfers in addition to the neutrino energy.

The matrix elements $\mathcal M_{\rm prod}(x,Q^2_p)$ 
and $\mathcal M_{\rm det}(y,Q^2_d)$
absorb the $Q^2$-dependent gauge couplings 
$g(Q^2_p)$ and $g(Q^2_d)$, respectively. 
For simplicity, we focus on the running effect
that enters the oscillation probability to emphasize
its effect on the oscillation behaviors.
However, the gauge coupling
can also 
enter the oscillation probability through 
the matter effect as we elaborate later in \gsec{sec:ME}. 

From \geqn{eq:amplitudeQFT},
an interesting effect emerges. The transition
probability does not
vanish in the zero-distance limit,
\begin{equation}
    P_{\alpha \beta} 
=
\left|
  \left[ U(Q_d^2) U^\dagger(Q_p^2) \right]_{\beta \alpha}
\right|^2,
\label{eq:Zero_distance_prob}
\end{equation}
if the mixing matrices $U(Q^2_p)$ and $U(Q^2_d)$ 
in the production and detection processes
mismatch with each other. On the other hand, the 
oscillation probability $P_{\alpha \beta}$
reduces to $\delta_{\alpha \beta}$ if $Q_d^2 = Q^2_p$. Notice that 
zero-distance represents the distance of a 
short-baseline experiment with $L \ll 2E_\nu/\Delta m_{ij}^2$. 
At the same time, the condition $|{\bf L}|\rightarrow\infty$ in \geqn{eq:GStheorem}
is still true as long as $L E_\nu \gg 1$ \cite{Grimus:1996av}. 
In other words, the so-called zero-distance is 
macroscopic and the neutrinos are on-shell, 
which means the production and detection processes are 
still separated. One should keep in mind that this is 
just a limit and not really zero distance.
Therefore, the Grimus-Stockinger theorem can 
still be used in the zero-distance limit 
since the condition $1/E_\nu \ll L \ll 2E_\nu/\Delta m_{ij}^2$ 
is valid for all the short-baseline neutrino 
experiments we consider later. 

The zero-distance transition probability 
can be explicitly calculated by using the
standard PMNS parametrization \cite{PDG22-NuCP},
\begin{equation}
    U_{\rm PMNS} 
\equiv
  O_{23} P_{\delta_D}^* O_{13} P_{\delta_D} O_{12},
\end{equation}
where $O_{ij}$ are the Euler rotation matrices
and $P_{\delta_D} \equiv {\rm diag} \{1,1, e^{-i\delta_D}\}$. 
Since we mainly focus on the RG running of 
the Dirac CP phase $\delta_D$, we simply fix 
the other oscillation parameters except the
CP phase and define the change of $\delta_D$ 
due to RG running as 
$\Delta \delta_D (Q^2_{d,p}) \equiv \delta_D (Q^2_d)  - \delta_D (Q^2_p)$, 
where $\delta_{p,d} \equiv \delta_D(Q^2_{p,d})$ represent
the running CP phases in production and 
detection, respectively. Then
the transition amplitude 
$U\left(Q_{d}^2\right)
U^\dagger\left(Q_{p}^2\right)$ becomes,
\begin{eqnarray}
  \mathbb 1 
+
 \frac{e^{-i \Delta \delta_D} - 1}{2}
\begin{pmatrix}
  2 \sin^2 {\theta_r} 
& e^{-i\delta_p} s_a \sin2\theta_r  
& e^{-i\delta_p} c_a \sin2\theta_r
\\ 
  e^{i\delta_d} s_a \sin2\theta_r 
&-2 e^{i \Delta \delta_D} s_a^2   \sin^2 \theta_r 
&-2 e^{i \Delta \delta_D} s_a c_a \sin^2 \theta_r 
\\ 
  e^{i\delta_d} c_a \sin2\theta_r 
&-2e^{i \Delta \delta_D} s_a c_a \sin^2 \theta_r 
&-2e^{i \Delta \delta_D} c_a^2   \sin^2 \theta_r
\end{pmatrix}. 
\label{eq:zero-distance-amplitude}
\end{eqnarray}
It is interesting to see that it contains only
the reactor ($\theta_r$) and atmospheric ($\theta_a$)
mixing angles while the solar angle ($\theta_s$) does not
appear at all.
Then the explicit form 
of the zero-distance transition probability
is obtained by putting the 
above expression into \geqn{eq:Zero_distance_prob}.
For $\nu_e \rightarrow \nu_e$
and $\nu_\mu \rightarrow \nu_e$, the transition
probabilities are,
\begin{eqnarray}
    P_{ee} (Q^2_{d,p}) 
=
  1 - \sin^2\left(\frac{\Delta \delta_D}{2}\right)\sin^2 2\theta_{13},
 \quad 
  P_{\mu e} (Q^2_{d,p}) 
=
  \sin^2\left(\frac{\Delta \delta_D}{2}\right)s^2_{23}\sin^2 2\theta_{13} \,.
  \label{eq:sblEQ}
\end{eqnarray}
Although these probabilities are independent of the 
Dirac CP phase itself, but are functions of
the CP phase difference induced by the 
running effect. Most interestingly, the $P_{ee}$
oscillation probability that is typically independent
of the Dirac CP phase has dependence on the CP phase
difference induced by the RG running parameter
$\Delta \delta_D$.
The zero-distance effect is unique for identifying
the running effect. On the
experimental side, such an effect can be 
constrained using the data from the 
short-baseline (SBL) experiments as elaborated 
later in \gsec{sec:SBLconstraint}.

\subsection{Neutrino Oscillation in Matter with RG Running}
\label{sec:ME}

The neutrinos propagating in matter can experience 
that matter effect that arises from the forward 
scattering mediated by weak gauge bosons. 
It adds an additional matter 
potential matrix $V$ to the vacuum Hamiltonian $H_0$,
\begin{align}
  H
\equiv
  H_0
+ V.
\end{align}
Note that the vacuum Hamiltonian
$H_0\equiv M^\dagger M/2E_\nu$ is usually defined 
in the vacuum flavor basis 
$\nu_\alpha (Q^2_{\rm vac})$ where the
momentum transfer is denoted as $Q^2_{\rm vac}$.
Although there is no unique definition 
of $Q^2_{\rm vac}$ and its value is unknown, 
the introduction of this momentum transfer
scale will not bring any problems since it will not
appear in the final expression of oscillation 
probability as we show below.
To ensure that the summation between $H_0$ 
and the matter potential matrix $V$ makes sense,
the latter needs to be written 
in the same flavor basis as $H_0$. 
\begin{figure}[t]
    \centering
    \includegraphics[scale=0.7]{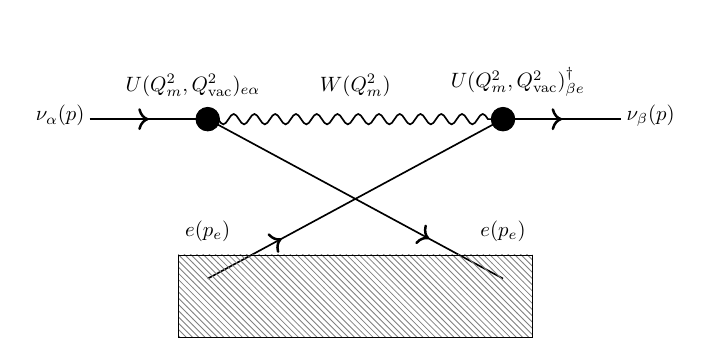}
    \includegraphics[scale=0.7]    {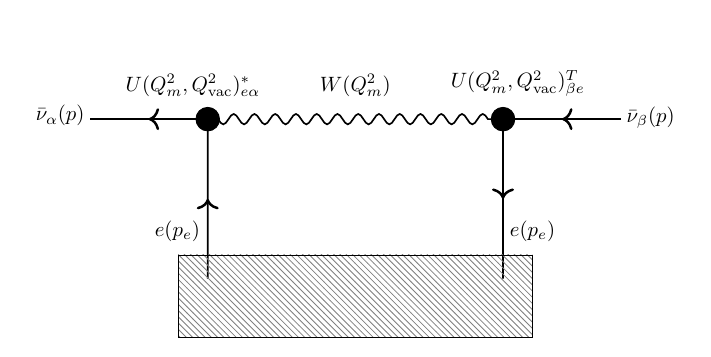}
    \caption{The Feynman diagrams for the neutrino 
    (left) and anti-neutrino (right) matter effects.
    }
    \label{fig:QFT_nuOSc_matter2}
\end{figure}

Recall that the neutrino flavor eigen-fields 
depend on the momentum transfer.
Similarly the matter flavor 
basis is defined at the 
momentum transfer $Q^2_m$
of neutrino-matter interactions. 
As specifically shown in \gfig{fig:QFT_nuOSc_matter2}, the 
neutrino and anti-neutrino matter interactions are 
contributed by different Feynman diagrams. 
Although both arise from the $W$ boson
mediation, their momentum transfers are different,
\begin{align}
  \nu: Q^2_m \equiv |(p_\nu - p_e)^2| = |m^2_e - 2 m_e E_\nu|,
\qquad
  \bar \nu: Q^2_m \equiv |(p_\nu + p_e)^2| = |m^2_e + 2 m_e E_\nu|,
\label{eq:Q2m}
\end{align}
which should not be a
vanishing value as adopted in
\cite{Babu:2021cxe}.
Between the neutrino and anti-neutrino modes, the momentum
transfer $Q^2_m = |m^2_e \mp 2 m_e E_\nu|$ differs.
Consequently, the matter flavor eigen-fields
with $Q^2_m$ dependence are no longer the same as 
their vacuum counterparts and the usual definition of
the matter potential without considering RG running
no longer holds.

The matter potential
matrix should be diagonal in the matter 
flavor basis $\nu_\alpha(Q^2_m)$,
\begin{align}
  V
\equiv
\begin{pmatrix}
  \sqrt 2 G_F n_e & 0 & 0 \\
  0 & 0 & 0 \\
  0 & 0 & 0
\end{pmatrix}.
\label{eq:H0V}
\end{align}
To match with the vacuum Hamiltonian $H_0$, the 
matter potential 
matrix needs to be rotated to the vacuum  
flavor basis $\nu_\alpha(Q^2_{\rm vac})$.
The connection between the vacuum 
($\nu_\alpha(Q^2_{\rm vac})$)
and matter ($\nu_\alpha(Q^2_m)$) 
flavor eigen-fields is established through 
the commonly shared mass eigen-fields $\nu_i$, 
$\nu_\alpha(Q^2) = \sum_i
U_{\alpha i}(Q^2) \nu_i$, 
\begin{eqnarray}
  \nu_\alpha (Q^2_m)
= \sum_i U_{\alpha i}(Q^2_m) \nu_i
=
  \sum_{i \beta} 
  U_{\alpha i}(Q^2_m)
  U_{\beta i}^\ast (Q^2_{\rm vac})
  \nu_\beta(Q^2_{\rm vac}).
\label{eq:VaMaFlavor}
\end{eqnarray}
For convenience, we define $
U(Q^2_m, Q^2_{\rm vac})
\equiv U(Q^2_m)
U^\dagger (Q^2_{\rm vac})$.
Although the production,
detection, and even the matter interaction
processes involve different momentum transfers 
and hence should have different flavor eigen-fields, 
the vacuum mass eigen-fields is always defined for 
on-shell neutrinos. This property
is based on the observation that the neutrino mass
receives a very tiny RG running effect \cite{Casas:1999tg,Chankowski:2001mx,Antusch:2003kp,Ray:2010rz,Ohlsson:2013xva,Babu:2021cxe}. 
Now the total Hamiltonian can be written in the
common vacuum flavor eigen-fields 
($\nu_\alpha(Q^2_{\rm vac})$)
using the transformation between vacuum and matter 
flavor eigen-fields in \geqn{eq:VaMaFlavor},
$  H
=
  M^\dagger M/2 E_\nu
  +
   U^\dagger(Q^2_m, Q^2_{\rm vac})
   V
   U(Q^2_m, Q^2_{\rm vac})$.
After a similarity transformation 
$H\rightarrow U^\dagger(Q^2_{\rm vac})HU(Q^2_{\rm vac})$, 
the total Hamiltonian can be written in the vacuum 
mass eigen-fields, 
\begin{align}
  H
=
  \frac {M_d^2}{2 E_\nu}+U^\dagger(Q^2_m)VU(Q^2_m), 
  \label{eq:MdV}
\end{align}
where $M_d^2 \equiv U^\dagger(Q^2_{\rm vac})M^\dagger M U(Q^2_{\rm vac})
\equiv{\rm diag} \{m^2_1, m^2_2, m^2_3\}$ 
is the diagonal mass matrix in vacuum. The 
undetermined $Q^2_{\rm vac}$ does not appear
in this Hamiltonian anymore and will not affect the 
oscillation probability.

The diagonalization of the total Hamiltonian 
will give the effective mass eigen-fields $\tilde \nu_i$ 
with the effective masses $\widetilde m_i$
during the propagation by one unitary matrix $U(Q^2_m,V)$, 
\begin{eqnarray}
  H_{d}
\equiv
  U^\dagger(Q^2_m, V) H U(Q^2_m, V)
\equiv 
  {\rm diag}
\left\{
  \frac {\widetilde m_1^2} {2 E_\nu},
  \frac {\widetilde m_2^2} {2 E_\nu},
  \frac {\widetilde m_3^2} {2 E_\nu}
\right\}.
\label{eq:diagH}
\end{eqnarray}
The vacuum mass eigen-fields 
$\nu_i$ and the effective mass 
eigen-fields $\tilde \nu_i$ in matter
are connected via the same unitary 
matrix $U(Q^2_m,V)$, $\nu_i=\sum_j U_{ij}(Q^2_m,V)\tilde\nu_j(Q^2_m,V)$.
%$\tilde \nu_i=\sum_j U^\ast_{ji}(Q^2_m,V)\nu_j$. 
It can reduce to $\nu_i = \widetilde \nu_i$ 
in the absence of matter potential $V$ as 
expected.
Note that the matrix $U(Q^2_m,V)$ is only 
used to diagonalize the total Hamiltonian.

To connect propagation with
the production and detection processes,
the relation between the effective mass 
eigen-fields
$\widetilde \nu_i$ in matter 
and the production as well as detection 
flavor eigen-fields $\nu_\alpha(Q^2_{p,d})$ 
is needed.
Since the production and detection 
flavor eigen-fields connect with 
the vacuum mass eigen-fields $\nu_j$ via
$\nu_\alpha(Q^2_{p,d}) =\sum_{i} U_{\alpha i} 
(Q^2_{p,d}) \nu_i$,
the effective mass eigen-fields 
correlate with the production and detection 
flavor eigen-fields,
\begin{align}
  \nu_\alpha(Q^2_{p,d})
=
  \sum_{ij} U_{\alpha i}(Q^2_{p,d})
  U_{ij}(Q^2_m,V)\tilde\nu_j(Q^2_m,V)
\equiv 
  \sum_\alpha U_{\alpha j}(Q^2_{p,d},Q^2_m,V)
  \tilde\nu_j(Q^2_m,V),
\label{eq:UQQV_def2}
\end{align}
where we define a unitary matrix 
$ U(Q^2_{p,d},Q^2_m,V)\equiv U(Q^2_{p,d})U(Q^2_m,V)$ for convenience.

%\begin{figure}[H]
%    \centering
%    \includegraphics[scale=0.6]{Figures/OscillationDiagram_matter.pdf}
%    \caption{QFT description of neutrino oscillation in matter.}
%    \label{fig:QFT_nuOSc_matter}
%\end{figure}
%

Now the oscillation formula for the vacuum case 
in \gsec{sec:vac-zd}
can easily extend to the matter case.
Reviewing the whole oscillation process shown in \gfig{fig:QFT_nuOSc},
the neutrino is first produced with 
flavor $\alpha$, which decomposes into effective
neutrino mass eigen-fields with an
associated matrix $U^\ast_{\alpha i}(Q^2_p,Q^2_m,V)$.
After propagating a macroscopic spatial distance $L$, each effective mass
eigen-field with the 
effective mass $\widetilde m_i$ 
receives a phase factor, $e^{- iL \frac {\widetilde m^2_i}{2 E_\nu}}$. 
Finally, the effective mass eigen-fields 
combine linearly with coefficients 
$U_{\beta i}(Q^2_d,Q^2_m,V)$ 
to flavor eigen-fields $\nu_\beta(Q^2_d)$ to be detected.
Then the net effect gives the oscillation amplitude,
\begin{eqnarray}
   \mathcal A_{\beta \alpha}^{\rm osc}
\equiv 
   \sum_{i}
    U_{\beta i}(Q^2_{d}, Q^2_m, V)
    e^{- iL \frac {\widetilde m^2_i}{2 E_\nu}}
    U_{i \alpha}^\dagger(Q^2_{p}, Q^2_m, V)
    \,. 
    \label{eq:amplitude-matter-QFT}
\end{eqnarray}
The above amplitude
reduces to the vacuum one when the matter potential 
is vanishing with 
$U(Q^2_{p,d},Q^2_m,V)\rightarrow U(Q^2_{p,d})$ 
and $\tilde m_i\rightarrow m_i$ 
as expected. Different from the
vacuum case, the oscillation amplitude with 
matter effect involves one more momentum transfer 
$Q^2_m$ which is not negligible when considering 
the RG running effect. Besides, the gauge coupling enters the oscillation amplitude/probability 
through the matter potential 
term of the total Hamiltonian
as shown in \geqn{eq:MdV}. 
For simplicity, we do not
consider the RG running 
effect on the coupling 
constant but only the mixing parameters in this paper.

\subsection{Numerical Illustration of the RG Running Effect}
\label{sec:numericalProb}

To illustrate the RG 
running effect on the neutrino
oscillation probabilities (dashed, dotted, and dash-dotted curves), 
\begin{figure}[t!]
\centering
\includegraphics[width=0.49\textwidth]{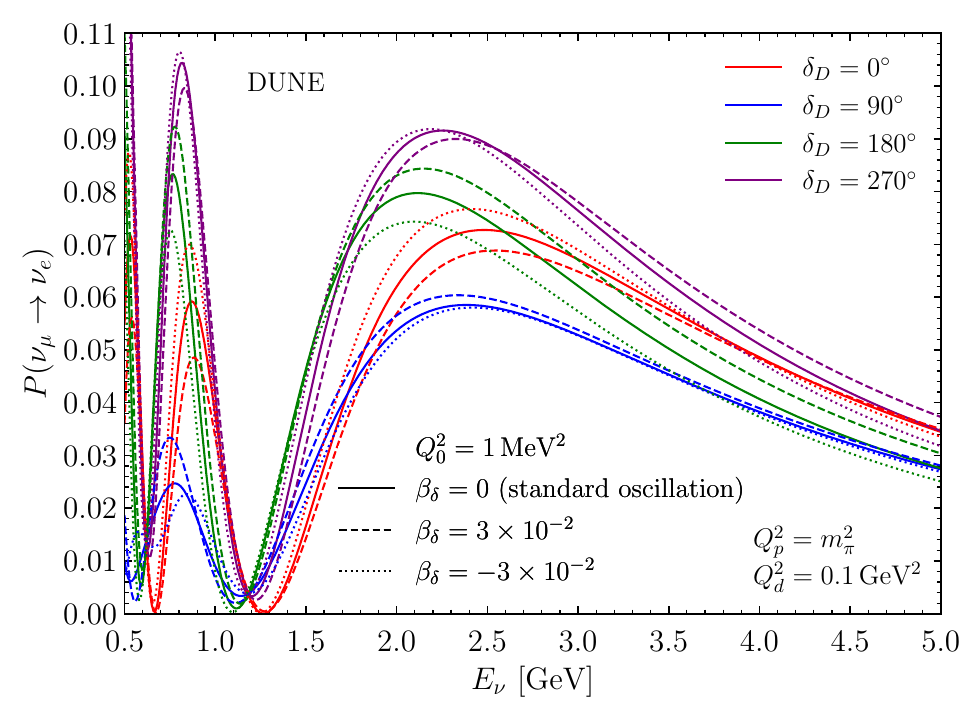}
\includegraphics[width=0.49\textwidth]{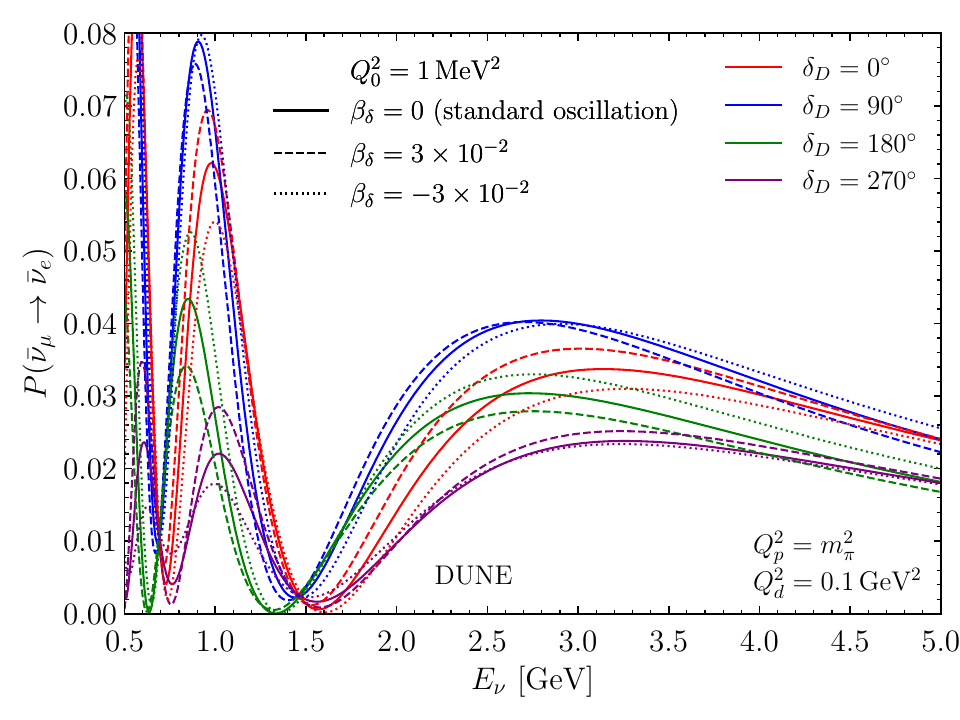}\\
\includegraphics[width=0.49\textwidth]{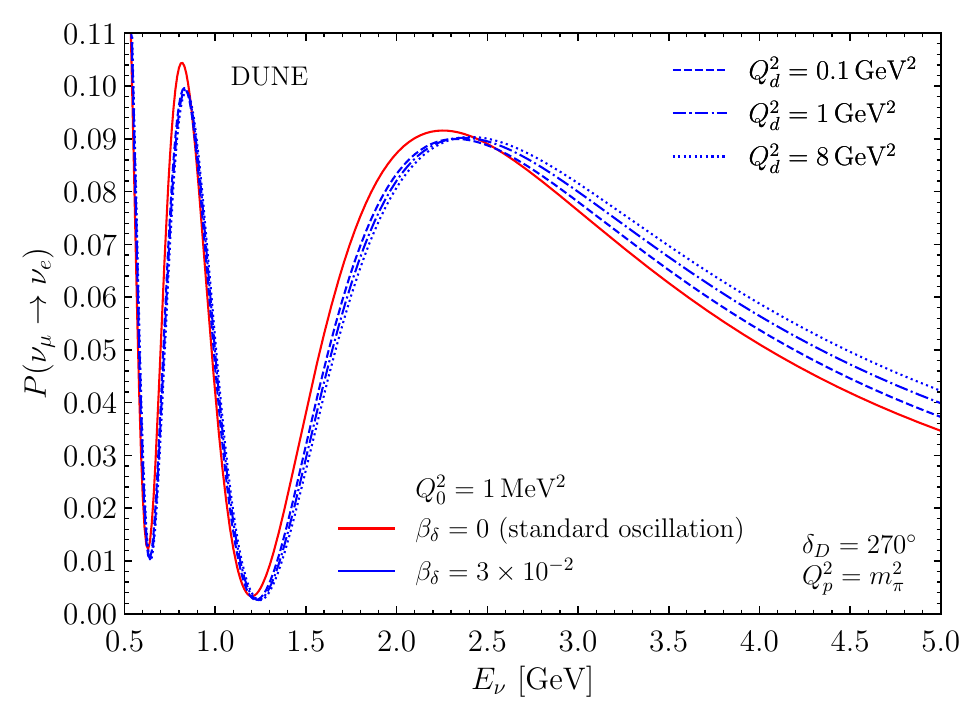}
\includegraphics[width=0.49\textwidth]{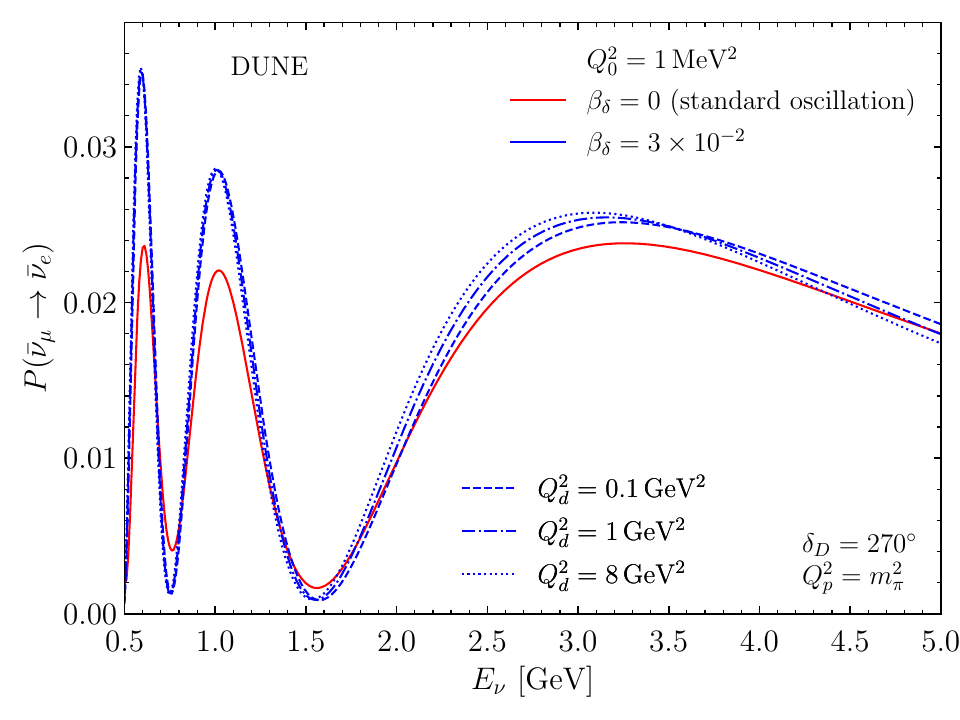}
\caption{The oscillation 
probabilities as a function of 
the neutrino energy $E_\nu$ for the DUNE 
configuration with 
$L=1284.9\,$km and $\rho = 
2.848\,$g/cm$^3$.
\textbf{Upper left:}  
The $\nu_\mu\rightarrow\nu_e$ 
oscillation 
probabilities 
for the neutrino mode with 
$\beta_\delta=0$ (solid), 
$3\times 10^{-2}$ (dashed), 
and $-3\times 10^{-2}$ 
(dotted). Note that
$\beta_\delta=0$ corresponds 
to the standard oscillation 
probabilities without RG 
running effect. 
Four different CP values 
$\delta_D=0^\circ$ (red), 
$90^\circ$ (blue), $180^\circ$ (green), and $270^\circ$ (purple) are chosen for illustration.
\textbf{Upper right:} Same as the left but for the 
anti-neutrino mode 
$\bar\nu_\mu\rightarrow\bar\nu_e$. 
\textbf{Lower left:} The 
$\nu_\mu\rightarrow\nu_e$ 
oscillation probabilities 
with fixed CP value 
$\delta_D=270^\circ$
for $Q^2_d=0.1\,$GeV$^2$ 
(dashed blue), $1\,$GeV$^2$ 
(dot-dashed blue), and 
8\,GeV$^2$ (blue dotted).  \textbf{Lower right:} Same as the left but for the 
anti-neutrino mode $\bar\nu_\mu\rightarrow\bar\nu_e$.
}
\label{fig:prob}
\end{figure}
we compare with the standard result
(solid curves) 
in \gfig{fig:prob}. 
For all panels, we take the DUNE configuration
with $L=1284.9\,$km, $E_\nu\in [0.5\,\rm{GeV},5\,\rm{GeV}]$, and production momentum transfer
$Q^2_p = (p_{\nu_\mu}+p_\mu)^2 = m^2_\pi$ for 
neutrinos produced from pion decay.
However, the detection momentum transfer 
$Q^2_d$ depends on the interaction
with the target and its value can vary 
a lot as we show 
below in \gsec{sec:experiment}.
The upper left/right panel shows
the RG running effect on the
$\nu_\mu\rightarrow\nu_e/\bar 
\nu_\mu\rightarrow\bar\nu_e$ 
oscillation probabilities
with three $\beta_\delta$ (defined in 
\geqn{eq:new_parametrization}) values: 
$\beta_\delta=0$ (without RG running), 
$-3\times 10^{-2}$, and 
$3\times 10^{-2}$. 
For both panels, there are three
oscillation peaks which locate at $E_\nu\sim
0.5$, $0.8$, $2.5\,$GeV, respectively. 
Among these three oscillation peaks, 
the 2.5\,GeV one contributes
most to the event rate at DUNE
since the DUNE neutrino 
spectrum peaks in this 
region. For neutrino energies 
within $[2\,\rm{GeV}, 
3\,\rm{GeV}]$, 
the relative oscillation
probability difference
($\equiv |P(\beta_\delta\neq 0)-P(\beta_\delta=0)|/P(\beta_\delta=0)$)
can reach percentage level
and has a similar size
as the CP effect. Hence 
the RG running effect on 
neutrino oscillation 
probabilities
is not negligible.

In the lower panels of \gfig{fig:prob}, we 
study the impact on oscillation
probabilities with three different $Q^2_d$ 
values: $Q^2_d=0.1\,$GeV$^2$, 1\,GeV$^2$, and 8\,GeV$^2$. 
For the neutrino mode,
the numerical result shows 
that the three blue
curves with different $Q^2_d$ values 
are close to each other for $E_\nu$ 
below $2.5\,$GeV
and can be distinguished for $E_\nu$ 
above $2.5\,$GeV.
However, for the anti-neutrino mode,
these curves are distinguishable for $E_\nu$
from 2\,GeV up to 5\,GeV.
Hence for a real neutrino experiment 
like DUNE, the momentum transfer should be
carefully reconstructed in order to 
study the RG running effect as we elaborate
below.

\section{CP Phase Sensitivity Reduction at DUNE}
\label{sec:4}

A major concern is that
the neutrino CP measurement will suffer from the 
RG running effect.
As mentioned above, 
the oscillation probability 
with RG running 
can vary a lot with the
neutrino energy $E_\nu$ and the
detection momentum transfer $Q^2_d$. 
Such effect appears when the two momentum
transfers in neutrino production ($Q^2_p$) and
detection ($Q^2_d$) mismatch. The larger
mismatch the larger RG running effect.
With large neutrino energy,
DUNE can provide a large
detection momentum transfer and 
hence a large RG running effect. 
So we take DUNE as an example to
quantitatively study the impact.

\subsection{Neutrino Energy and Momentum Transfer Reconstructions}
\label{sec:experiment}

The production momentum transfer has a fixed value 
as the pion mass $Q^2_p = m_\pi^2$ since 
most DUNE neutrinos  
are generated through pion decay \cite{DUNE:2020ypp}. 
To evaluate the oscillation probability 
$P_{\alpha \beta}(E_\nu, Q^2_p, Q^2_d)$ 
with the RG running effect, one still needs to reconstruct
the neutrino energy $E_\nu$ and the detection momentum 
transfer $Q^2_d$. 
The neutrino energy, $E^{\rm rec}_\nu \equiv 
E^{\rm rec}_\ell + E^{\rm rec}_{\rm had}$,
is reconstructed from the deposited
lepton ($E^{\rm rec}_\ell$) and hadronic
($E^{\rm rec}_{\rm had}$) energies \cite{DUNE:2015lol}.
On the other hand,
the reconstructed detection momentum 
transfer $Q^2_{\rm rec}$ 
correlates to not just the reconstructed 
neutrino energy $E^{\rm rec}_\nu$, 
but also the measured charged lepton energy 
$E^{\rm rec}_\ell$ and the measured 
lepton scattering angle
$\theta^{\rm rec}_\ell$,
\begin{eqnarray}
   Q^2_{\rm rec}
\equiv 
   |(p^{\rm rec}_\nu - p^{\rm rec}_\ell)^2|
= 
  2 E^{\rm rec}_\nu (E^{\rm rec}_\ell - |{\boldsymbol p}^{\rm rec}_\ell| \cos\theta^{\rm rec}_\ell)
- m^2_\ell,
\label{eq:Q2rec}
\end{eqnarray}
where $m_\ell$ is the lepton mass. 
Note that the lepton momentum cannot be directly 
reconstructed but needs to be obtained from the
reconstructed lepton energy, 
$|{\boldsymbol p}_\ell|^{\rm rec} \equiv \sqrt{(E^{\rm rec}_\ell)^2-m^2_\ell}$. 
In total, three observables (the reconstructed 
lepton energy $E^{\rm rec}_\ell$, the hadronic
energy $E^{\rm rec}_{\rm had}$, and the measured 
lepton scattering angle $\theta^{\rm rec}_\ell$)
are needed.

\subsubsection{Energy Reconstruction}

\begin{figure}[t!]
\centering
\includegraphics[width=0.49\textwidth]{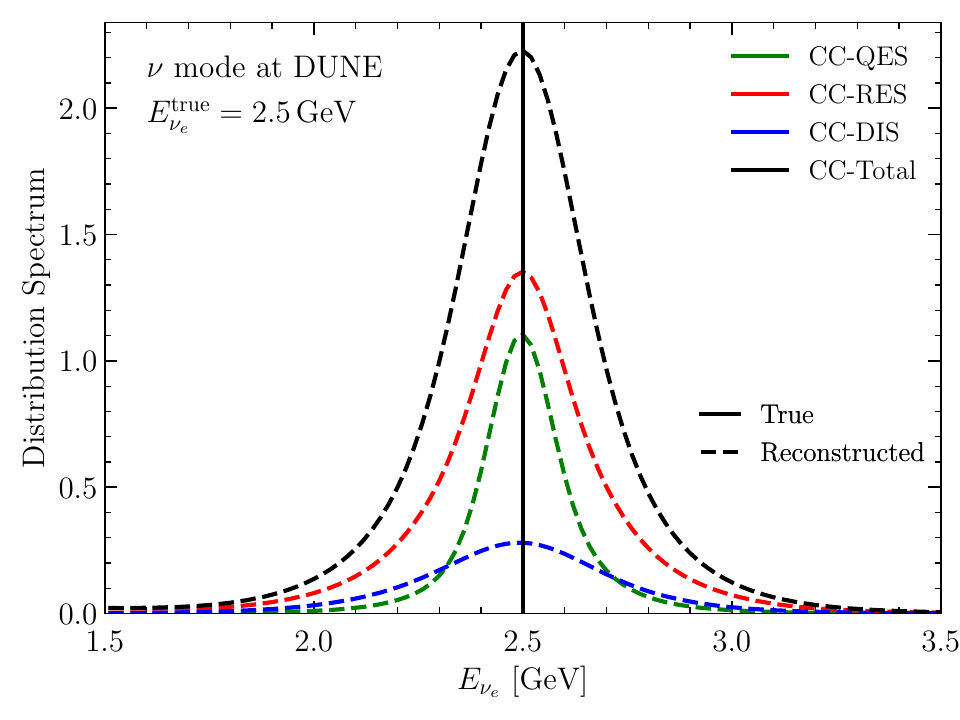}
\includegraphics[width=0.49\textwidth]{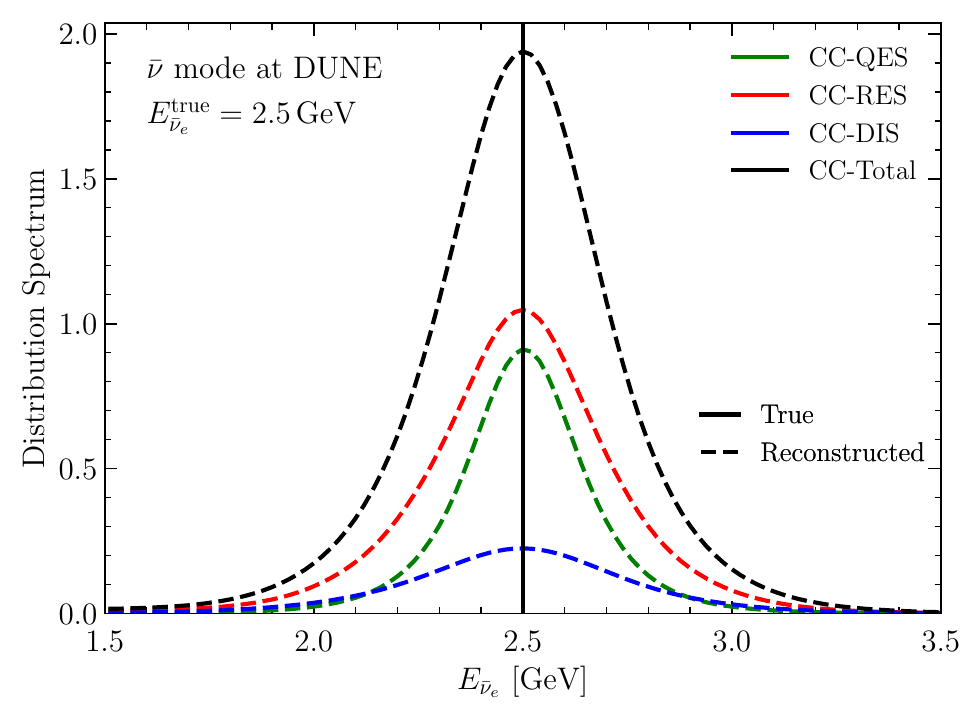}
\caption{
\textbf{Left:} The reconstructed energy ($E^{\rm rec}_\nu=
E^{\rm rec}_\ell+E^{\rm rec}_{\rm had}$) distribution
of a 2.5\,GeV $\nu_e$ neutrino at the DUNE
detector for 
different CC scattering processes: 
QES (green), RES (red), and DIS (blue). 
The combined total spectrum is plotted in black color. 
Note that the total spectrum is normalized and
the others are weighted by their corresponding cross sections.
\textbf{Right:} Same as the left panel but for the anti-neutrino mode.
}
\label{fig:Erec}
\end{figure}

We use GENIE \cite{Andreopoulos:2009rq,Andreopoulos:2015wxa} 
to generate the neutrino-Argon scattering 
events and simulate the neutrino energy reconstruction
at DUNE. Note that the neutrino energy does not 
fully deposit as visible 
energy and some is missing.  
In a liquid Argon Time Projection Chamber (TPC)
such as the DUNE far detectors, 
the energy is 
reconstructed via the
tracks left by the final-state particles. 
Those tracks with 
deposited energy below the 
detector threshold
cannot be measured. Moreover,
some final-state 
particles such as neutrino cannot
leave a track 
and hence is not measurable either.
These contribute as missing 
energy
\cite{Friedland:2018vry}.
To precisely reconstruct the neutrino energy, 
we add a 
constant value, $\langle E_{\rm  mis} \rangle$, as 
correction to calibrate the energy reconstruction
by taking into account the 
average missing energy, 
$\langle E_{\rm  mis} \rangle = 
30$\,MeV, $20$\,MeV, 
and $20$\,MeV for the 
QES, RES, and DIS 
scattering processes, respectively. 

In addition, the charged lepton
($E_\ell^{\rm rec}$) and hadronic
($E_{\rm had}^{\rm rec}$) energies are 
smeared around their 
respective true values. 
The Gaussain smearing resolution of 
each detectable particle is determined by
the \textit{best reconstruction case} 
in \cite{Friedland:2018vry} as summarized in 
\gtab{tab:resolution}. 
\begin{table}[h]
\centering
\begin{tabular}{c|c}
Particle &   Energy Resolution ($\sigma/E$) at DUNE 
\\
\hline
$\pi^{\pm}$ 
& 15\%
\\
$\mathrm{e}^{\pm} / \gamma$ 
& 1.5\%
\\
$p$ 
& 
\makecell{$\mathrm{p}<400\, \mathrm{MeV} / \mathrm{c}: 4 \%$\\ $\mathrm{p}>400\, \mathrm{MeV} / \mathrm{c}: 10 \% \oplus 3.1 \% / \sqrt{E}\,[\mathrm{GeV}]$}
\\
$n$
& $40 \% / \sqrt{E}\,[\mathrm{GeV}]$ 
\\
other 
& $5 \% \oplus 30 \% / \sqrt{E}\,[\mathrm{GeV}]$ 
\\
\end{tabular}
\caption{The energy resolutions for detectable particles 
obtained by fitting to the best reconstruction 
case in \cite{Friedland:2018vry}. }
\label{tab:resolution}
\end{table}

\gfig{fig:Erec} shows the reconstructed 
spectrum of CC events for 2.5\,GeV 
$\nu_e$ and $\bar\nu_e$, 
respectively. Note that the energy 
reconstruction for the CC-QES process
has a better performance than the other two
CC-RES and CC-DIS processes, since the 
deposited energy for CC-QES mostly comes
from the charged leptons which have 
a better resolution than the other particles 
produced in the RES and DIS processes.
\begin{figure}[t!]
\centering
\includegraphics[width=0.8\textwidth]{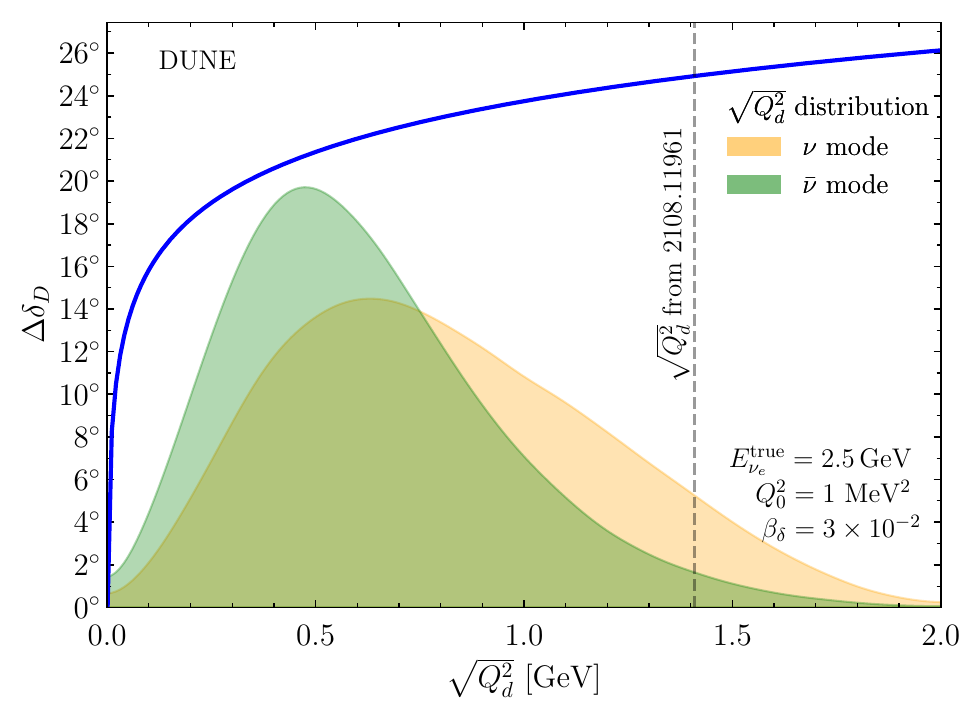}
\caption{
The CP phase change $\Delta\delta_D\equiv
\beta_\delta\ln\left(\left|Q^2/Q^2_0\right|\right)$ 
induced by the RG running effect with
$Q^2_0 = 1$\,MeV$^2$ and
$\beta_\delta = 3 \times 10^{-2}$ as function of
the detection momentum transfer $\sqrt{Q^2_d}$.
The $\sqrt{Q^2_d}$ distributions for
$\nu_e-$Argon scattering events
are shown with the orange and 
green shaded regions for the neutrino 
and anti-neutrino modes, respectively.
For comparison, the fixed
$\sqrt{Q^2_d}\approx 1.4\,$GeV
from $Q^2_d = 2m_NE_\nu^2/(2 
E_\nu + m_N)$
is presented by the vertical gray dashed line. 
}
\label{fig:scale}
\end{figure}

\subsubsection{Momentum Transfer Reconstruction in Terms of
Scattering Angle}

As mentioned above, the production momentum transfer 
$Q^2_p$
has fixed value as the pion mass, $Q^2_p = m_\pi^2$. 
For the detection momentum transfer $Q^2_d$,
the previous work \cite{Babu:2021cxe} also
uses a fixed value, $Q^2_d = 2m_NE_\nu^2/(2 
E_\nu + m_N)$,
where $m_N$ is the target nucleon mass. 
Then a 2.5\,GeV neutrino corresponds to 
$\sqrt{Q^2_d} \approx 1.4$\,GeV. 
However, the momentum transfer of 
CC events is not fixed but follows 
a distribution shown as the shaded 
regions in \gfig{fig:scale}. The 
estimated momentum transfer value actually
locates at the tail of the distribution and
is at least two times of the mean momentum 
transfer that is around $0.5 \sim 0.7$\,GeV. 
To see the running effect on the Dirac CP 
phase, we show the CP phase change 
$\Delta\delta_D$ $(\equiv 
\beta_\delta\ln\left(\left|Q^2/
Q^2_0\right|\right)$ 
induced by RG running with the
blue curve in \gfig{fig:scale}.
The value of $\Delta\delta_D$ varies a lot
throughout the $\sqrt{Q^2_d}$ distribution region.
Especially for $\sqrt{Q^2_d}<0.5\,$GeV,
the corresponding $\Delta\delta_D$ has 
a large variation and is quite different 
from its counterpart adopted in the previous work.
Therefore, an event-by-event reconstruction 
of the detection momentum transfer according
to \geqn{eq:Q2rec} is necessary at DUNE.

\begin{figure}[t!]
\centering
\includegraphics[scale = 1.05]{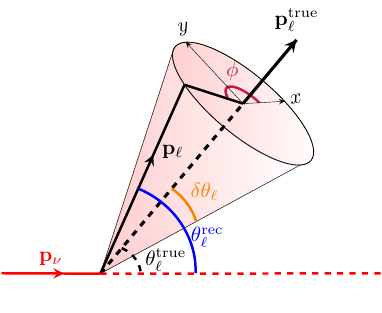}
\includegraphics[scale = 0.49]{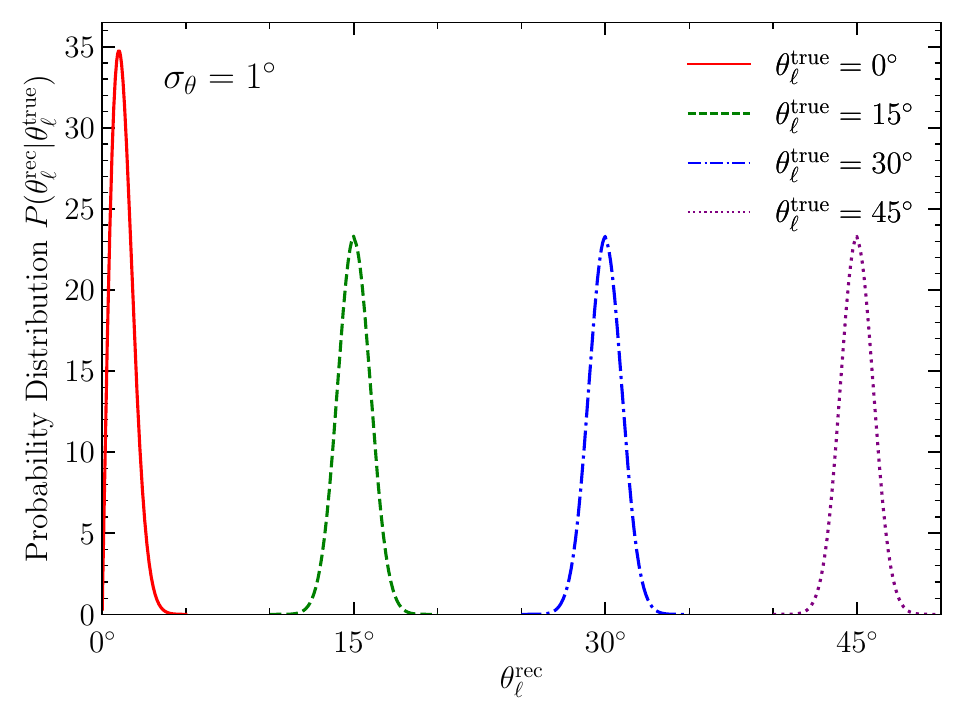}
\caption{
\textbf{Left:} The schematic plot for  
the charged lepton
scattering angle $\theta^{\rm rec}_\ell$ reconstruction. 
With $\theta_\ell^{\rm true}$ being the true 
scattering angle between the final-state
lepton ${\bf p}_\ell^{\rm true}$ and 
incoming neutrino ${\bf p}_\nu$ momenta, 
$\delta\theta_\ell$ is the 
opening angle between the measured lepton 
$\bf{p}_\ell$ and the true 
one ${\bf p}_\ell^{\rm true}$.
Moreover, the azimuthal angle 
$\phi$ for the measured lepton 
momentum surrounding the true lepton momentum
is randomly distributed in the range of [0, 2$\pi$).
\textbf{Right:}
The probability distribution of 
the lepton scattering angle $\theta^{\rm rec}_\ell$ with
four true scattering angles:  
$\theta^{\rm true}_\ell=0^\circ$ (solid red),
$15^\circ$ (dashed green), $30^\circ$ 
(dot-dashed blue), and $45^\circ$ (dotted 
purple). 
We use the  angular resolution 
$\sigma_\theta= 1^\circ$ 
from the DUNE CDR result.}
\label{fig:angular}
\end{figure}

In order to reconstruct the detection momentum
transfer $Q^2_d$, the scattering angle 
reconstruction plays an important role.
Note that the Gaussian 
smearing of the true
scattering angle cannot be directly 
used to simulate
the measured angle due to the geometry effect
that is illustrated in the left panel of \gfig{fig:angular}.
The incoming 
neutrino with momentum ${\bf p}_\nu$ 
interacts with the detector target which 
is marked as the black dot and 
a final-state charged lepton with 
true momentum ${\bf p}_\ell^{\rm true}$ 
is produced. Since the detector has a 
limited angular resolution, it measures
the lepton momentum ${\bf p}_\ell$ with 
an opening angle $\delta\theta_\ell$ from
the true one.
The value of $\delta\theta_\ell$ follows 
a Gaussian distribution that depends on
the detector angular resolution
$\sigma_\theta$. In addition,
the azimuthal angle $\phi$ for the measured 
lepton momentum surrounding the true lepton
momentum distributes randomly in [0, $2\pi$)
and affects the measured lepton scattering 
angle $\theta^{\rm rec}_\ell$. We follow the analytical 
method in \cite{Ge:2013ffa}
where the probability 
density $P(\theta^{\rm rec}_\ell|\theta_\ell^{\rm true})$ 
for obtaining the measured
scattering angle $\theta^{\rm rec}_\ell$  
from the given true one 
$\theta_\ell^{\rm true}$ is,
\begin{eqnarray}
    P(\theta^{\rm rec}_\ell|\theta_\ell^{\rm true})
=
  \frac{1}{2\pi} 
  \int^{\pi}_0 
  \frac{P(\delta\theta_\ell) \sin\theta^{\rm rec}_\ell}
  {
  \sqrt{\sin^2\theta^{\rm true}_\ell 
  \sin^2\delta\theta_\ell 
  -
  (\cos\theta^{\rm true}_\ell\cos\delta\theta_\ell
  - \cos\theta^{\rm rec}_\ell)^2}
  }
  d\delta\theta_\ell\,.
  \label{eq:thetarec-prob2}
\end{eqnarray}
In the above expression, 
$P(\delta\theta_\ell)\equiv 
\sin \delta\theta_\ell/N(\sigma_\theta)
\exp\left[-1/2\left(\delta\theta_\ell
/\sigma_\theta\right)^2\right]$ 
is the distribution function of 
the opening angle 
$\delta\theta_\ell$ with
$\sigma_\theta$ being the detector 
angular resolution and 
$N(\sigma_\theta)$ the
normalization factor.  
In our analysis, we take 
$\sigma_{\theta} = 1^\circ$ from 
the DUNE Conceptual Design Report (CDR) result 
\cite{DUNE:2015lol}. In the right panel of
\gfig{fig:angular}, we show the probability density 
for the measured scattering angle $\theta^{\rm rec}_\ell$
with four true scattering
angle values $\theta^{\rm true}_\ell=0^\circ$, $15^\circ$, $30^\circ$, 
and $45^\circ$ for illustration. For $\theta^{\rm true}_\ell\gg \sigma_\theta$, the
smeared scattering angle 
closely follows a Gaussian distribution
while the shape is significantly distorted 
when the $\theta^{\rm true}_\ell$ value is comparable
with its angular resolution $\sigma_\theta$.

\begin{figure}[t!]
\centering
\includegraphics[width=0.49\textwidth]{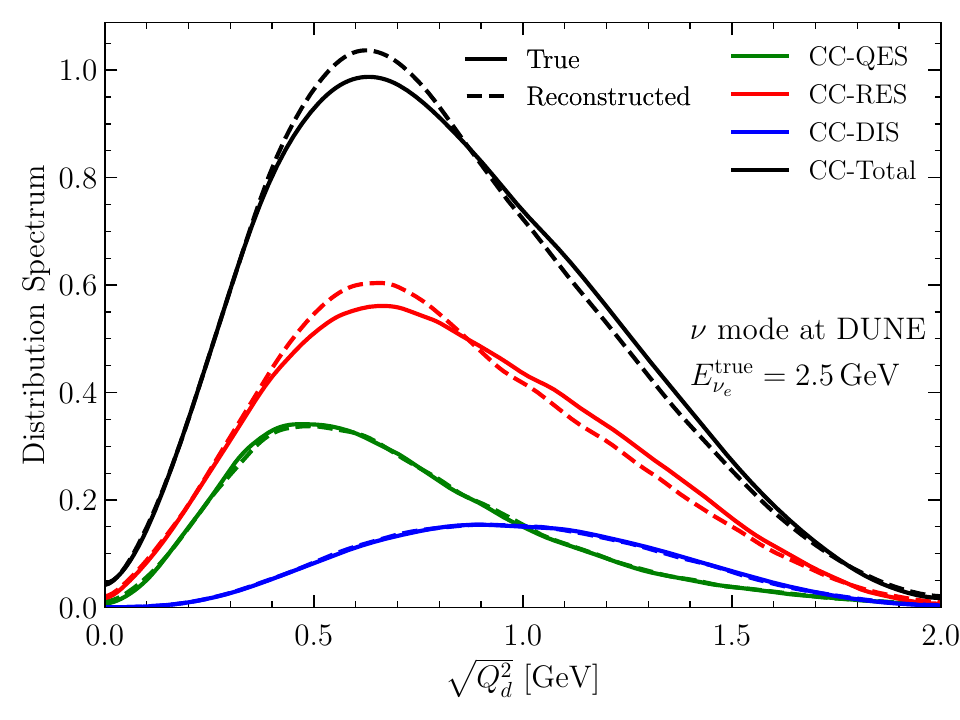}
\includegraphics[width=0.49\textwidth]{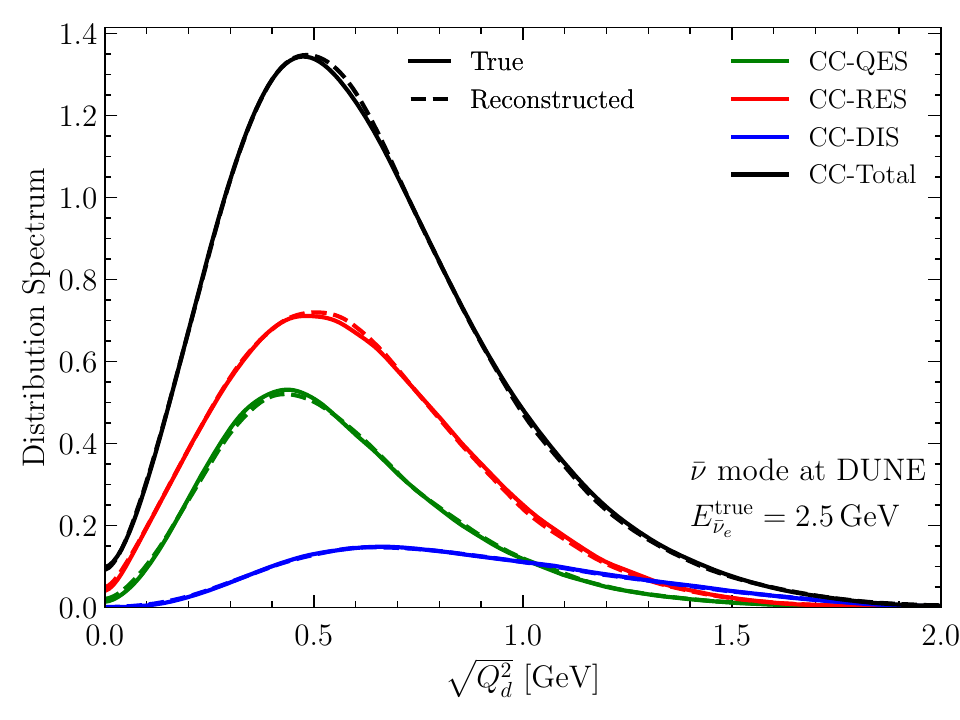}
\caption{
\textbf{Left:} The reconstructed momentum 
transfer $\sqrt{Q^2_d}$ distribution 
of a 2.5 GeV $\nu_e$ neutrino at the DUNE
detector for different CC scattering 
processes: QES (green), RES (red), and DIS 
(blue). The total spectrum (black) is normalized to 1 while
the others are weighted by their corresponding cross sections. 
\textbf{Right:} Same as the left panel
but for the anti-neutrino mode.
}
\label{fig:Qrec}
\end{figure}

Finally, the combination of 
$E^{\rm rec}_\nu, E^{\rm rec}_\ell$, and $\theta^{\rm rec}_\ell$ 
through \geqn{eq:Q2rec} reconstructs 
the detection momentum transfer 
$Q^2_d$.
\gfig{fig:Qrec} shows the 
$\sqrt{Q^2_d}$ distribution 
of a 2.5\,GeV neutrino scattering with 
Argon target for different CC scattering 
processes. The reconstructed momentum transfer 
distributions (dashed) are in good 
agreement with the true ones (solid), 
although the neutrino 
energy reconstruction is far from being perfect
as shown in \gfig{fig:Erec}.

\subsubsection{4-D Transfer Table for Energy and Momentum Transfer}
\label{sec:4-DTT}

As mentioned above, the detector energy and 
angular resolutions will
smear the true values $(E^{{\rm true},i}_\nu, Q^2_{{\rm true},j})$ 
into their 
reconstructed counterparts 
$(E_\nu^{{\rm rec},i'}, Q^2_{{\rm rec},j'})$ 
with $i$, $j$, $i'$, and $j'$
being the indices of their respective
bins.  
Given a true neutrino energy and a 
true momentum transfer, 
the transfer between the true and
reconstructed values is summarized in a
4-D transfer table $T'_{iji'j'}$,
\begin{eqnarray}
  T'_{iji'j'}
\equiv
  \frac{\Delta N(E^{{\rm true},i}_\nu, Q^2_{{\rm true},j}; E_\nu^{{\rm rec},i'}, Q^2_{{\rm rec},j'})}{N_{\rm total}(E^{{\rm true},i}_\nu, Q^2_{{\rm true},j})}.
\label{eq:original-transfer-table-element}
\end{eqnarray}
Here $\Delta N(E^{{\rm true},i}_\nu, Q^2_{{\rm true},j}; 
E_\nu^{{\rm rec},i'}, Q^2_{{\rm rec},j'})$ is the simulated event number 
within the $i'$-th 
reconstructed energy bin and
the $j'$-th reconstructed 
momentum transfer bin for the given
$i$-th true neutrino energy bin and
the $j$-th true momentum 
transfer bin,
while $N_{\rm total}(E^{{\rm true},i}_\nu, Q^2_{{\rm true},j})$ 
is the total simulated event 
number of the given $i$-th true
neutrino energy bin and $j$-th 
true momentum transfer bin. 
The transfer table element
is the probability 
for finding the reconstructed 
values ($E_\nu^{{\rm rec},i'}, Q^2_{{\rm rec},j'}$) 
after smearing their respective true
values ($E^{{\rm true},i}_\nu, Q^2_{{\rm true},j}$). 
Such effective
transfer table satisfies the probability 
conservation condition, $\sum_{i'j'}T'_{iji'j'}=1$, as expected.
Moreover, the transfer table enters 
the reconstructed event
rates calculation 
as elaborated below.

\subsection{Extended GLoBES Simulation with both Energy and Momentum Transfer}

As pointed out around \geqn{eq:amplitudeQFT},
the oscillation probability has momentum transfer
dependence. In the presence of the RG running effect,
the neutrino oscillation event rate is a function
of not just the neutrino energy ($E_\nu$) 
but also the
momentum transfers ($Q^2_p$ and $Q^2_d$). With 
the production momentum transfer ($Q^2_p$)
typically fixed by 
the parent meson masses, the event rate
is still a two-dimensional distribution of
the neutrino energy ($E_\nu$) 
and the detection momentum transfer ($Q^2_d$).

The GLoBES \cite{Huber:2004ka,Huber:2007ji} simulation for fixed 
baseline experiments with one-dimensional 
event rate needs to be extended in order
to incorporate the RG running effect.
The two-dimensional binning approach 
with neutrino energy and zenith angle (baseline 
length) dependence
has been used in the atmospheric neutrino 
oscillation analysis \cite{Winter:2013ema,Winter:2015zwx}. 
We adopt similar approach by breaking 
the 2-D event rate $d^2 N / d E_\nu d Q^2_d$ into
multiple \texttt{glb} experimental 
profiles each with a fixed $Q^2_d$ and 
merging them in the end. 

\begin{itemize}[leftmargin=*]
\item {\bf GLoBES simulation}: With a fixed $Q^2_d$ for each \texttt{glb} profile,
the oscillation probability
$P_{\alpha \beta}(E_{\nu}^{{\rm true},i}, Q^2_{{\rm true},j})$
with the RG running effect in
\gsec{sec:ME} taken into account and the corresponding event 
spectrum can be calculated in the usual way by GLoBES,
\begin{eqnarray}
  \frac{dN^{\rm true}}{d E_{\nu}^{{\rm true},i}} \Bigg|_{Q^2_{{\rm true},j}}
\equiv 
  N^{\rm target} 
  \phi_\alpha(E_{\nu}^{{\rm true},i}) 
  P_{\alpha \beta}(E_{\nu}^{{\rm true},i}, Q^2_{{\rm true},j}) 
  \sigma_\beta(E_{\nu}^{{\rm true},i}),
  \label{eq:GLB-j}
\end{eqnarray}
where $N^{\rm target}$ is the normalization factor including
the information of running time, detector size, and 
the baseline length.
Both the $\alpha$-flavor neutrino flux 
$\phi_\alpha(E_{\nu}^{{\rm true},i})$ and $\beta$-flavor
detection cross section $\sigma_\beta(E_{\nu}^{{\rm true},i})$
are functions of the true neutrino energy
$E_{\nu}^{{\rm true},i}$ for the $i$-th bin.

\hspace{3mm}
Since the neutrino energy $E_\nu$ and detection momentum transfer
$Q^2_d$ are reconstructed altogether, their smearing
interleaves with each other and cannot be separated.
So we first extract from GLoBES the event numbers at
the true neutrino energy level,
\begin{eqnarray}
  {\rm GLB}_j 
 :\quad 
  N^{{\rm GLB}_j}_i
=
  \frac{dN^{\rm true}}{d E_{\nu}^{{\rm true},i}} \Bigg|_{Q^2_{{\rm true},j}}
  \Delta E_i^{{\rm true},i},
\label{eq:Ni}
\end{eqnarray}
for the $j$-th \texttt{glb} experimental profile
without smearing.

\item {\bf Implementing the momentum transfer distribution by hand}:
Although GLoBES automatically 
incorporates the total cross section 
$\sigma(E_{\nu}^{{\rm true},i})$ 
in the event rate calculation according to
\geqn{eq:GLB-j}, the differential spectrum
with respect to the momentum transfer $Q^2_d$
is still missing. One needs to multiply the
extracted event rate $N^{{\rm GLB}_j}_i$ with
a normalized differential cross section
$d\sigma_\beta/(\sigma_\beta dQ^2_{{\rm true},j})$
by hand. The event numbers among different
\texttt{glb} profiles are then reshuffled into,
\begin{eqnarray}
  N_{ij}
\equiv
  N^{{\rm GLB}_j}_i 
  \times
  \left(
  \frac{d\sigma_\beta}{\sigma_\beta dQ^2_{{\rm true},j}}
  \right)\Bigg|_{E^{{\rm true},i}_\nu}
  \Delta Q^2_{{\rm true},j},
  \label{eq:EN1} 
\end{eqnarray}
weighted by the momentum transfer bin size 
$\Delta Q^2_{{\rm true},j}$. The product 
of the normalized differential cross
section and the true momentum transfer bin 
size can be simulated by GENIE as,
$d \sigma_\beta /(\sigma_\beta d Q^2_{{\rm true},j})|_{E^{{\rm true},i}_\nu} 
\Delta Q^2_{{\rm true},j}=\Delta N(E^{{\rm true},i}_\nu, Q^2_{{\rm true},j})/N(E^{{\rm true},i}_\nu)$, 
as a ratio between the event number 
within the $j$-th true momentum transfer 
bin and the total one
for a given true energy.

\hspace{3mm}
Note that the two variables $N^{{\rm GLB}_j}_i$ and $N_{ij}$
have different physical meanings. While
$N^{{\rm GLB}_j}_{i}$ is the event number within the 
$i$-th energy bin for the $j$-th \texttt{glb} profile
that carries the momentum transfer dependence
in the oscillation probability but no information
about the momentum transfer distribution,
the event number
$N_{ij}$ within the $i$-th energy 
and $j$-th momentum transfer bin also includes the differential momentum transfer distribution.
The essential difference between these two variables 
is having the momentum transfer distribution or not.
Nevertheless, the $j$ index of $N^{{\rm GLB}_j}_{i}$ cannot be
simply omitted since the oscillation probability has momentum
transfer dependence. To make this tricky point explicit,
we add $j$ as an upper index for $N^{{\rm GLB}_j}_{i}$ and
a lower one for $N_{ij}$.

\item {\bf Detector resolution and smearing}:
The normalized differential 
cross section in the 2-D
true event rate calculation \geqn{eq:EN1}
cannot be automatically 
carried out in GLoBES 
and needs to be implemented by hand.
In addition, the 4-D transfer table $T'_{iji'j'}$ 
defined in \geqn{eq:original-transfer-table-element}
also needs to be implemented 
by hand with multiple \texttt{glb} 
profiles.
For convenience, we combine these
two terms into a single effective 
4-D transfer table, 
\begin{eqnarray}
  N_{i'j'} 
\equiv
  \sum_{ij}
  N_{ij}
\times 
  T'_{iji'j'}
%\times 
%  \Delta E^{{\rm rec},i'}_\nu
%\times
%  \Delta Q^2_{{\rm rec},j'}
\label{eq:ENrec1}
\equiv
  \sum_{ij}
  N^{{\rm GLB}_j}_i 
\times
  T_{iji'j'},
\label{eq:ENrec}
\end{eqnarray}
where $N_{i'j'}$ is the event number 
in the $i'$-th reconstructed energy 
and $j'$-th reconstructed
momentum transfer bin. 
Instead of using the $T'_{iji'j'}$
\geqn{eq:original-transfer-table-element},
the event rate $N^{{\rm GLB}_j}_i$
extracted from GLoBES is directly 
related to the reconstructed event rate
$N_{i'j'}$ with, 
\begin{align}
  T_{iji'j'}
\equiv
  \left(
  \frac{d\sigma_\beta}{\sigma_\beta dQ^2_{{\rm true},j}}
  \right)\Bigg|_{E^{{\rm true},i}_\nu}
\times
  \Delta Q^2_{{\rm true},j}
\times
  T'_{iji'j'}
=
  \frac{\Delta N(E^{{\rm true},i}_\nu, Q^2_{{\rm true},j}; E_\nu^{{\rm rec},i'}, Q^2_{{\rm rec},j'})}{N_{\rm total}(E^{{\rm true},i}_\nu)},
%\times
%  \Delta E^{{\rm rec},i'}_\nu
%\times
%  \Delta Q^2_{{\rm rec},j'},
\label{eq:TransferTab}
\end{align}
where $N_{\rm total}(E^{{\rm true},i}_\nu)$ 
is the total event number for 
the given $i$-th true
neutrino energy bin. 
The effective transfer table
includes three terms: the normalized differential cross section 
$d \sigma_\beta /(\sigma_\beta d Q^2_{{\rm true},j})|_{E^{{\rm true},i}_\nu}$, the bin 
size $\Delta Q^2_{{\rm true},j}$, 
and the smearing transfer table $T'_{iji'j'}$, which can be combined
into a single simulation. 

\hspace{3mm}
In our simulation, we first use GENIE to 
generate $10^6$ neutrino-Ar
scattering events 
for the true energy $E^{{\rm true},i}_\nu$ 
at the center of each bin. 
We then reconstruct event by event
the neutrino energy and 
momentum transfer
after smearing according to 
the detector energy and
angular resolutions as discussed in \gsec{sec:experiment}. 
For the true/reconstructed energy, 
we take a 0.25\,GeV bin size 
from 0.5 to 5\,GeV. 
On the other hand,
the true/reconstructed momentum transfer
uses $\sqrt{Q^2_d}$ instead of 
$Q^2_d$ with 25 bins from 0 to 
$2\sqrt{2}\,$GeV 
since $Q^2_d$ 
distributes mainly at small 
values as shown in \gfig{fig:Qrec}. 
Finally, we bin the events 
into a histogram with
variables $(E_\nu^{\rm true}, Q^2_{\rm true}; E_\nu^{\rm rec}, Q^2_{\rm rec})$ 
and divide the event number in each bin by
the total event number to obtain the
$T_{iji'j'}$ defined in 
\geqn{eq:TransferTab}.
After these, we calculate the
2-D event rates according to 
\geqn{eq:ENrec} and feed them 
back into GLoBES for $\chi^2$
analysis to take full advantage. 
\end{itemize}

\begin{figure}[t!]
\centering
\includegraphics[width=0.49\textwidth]{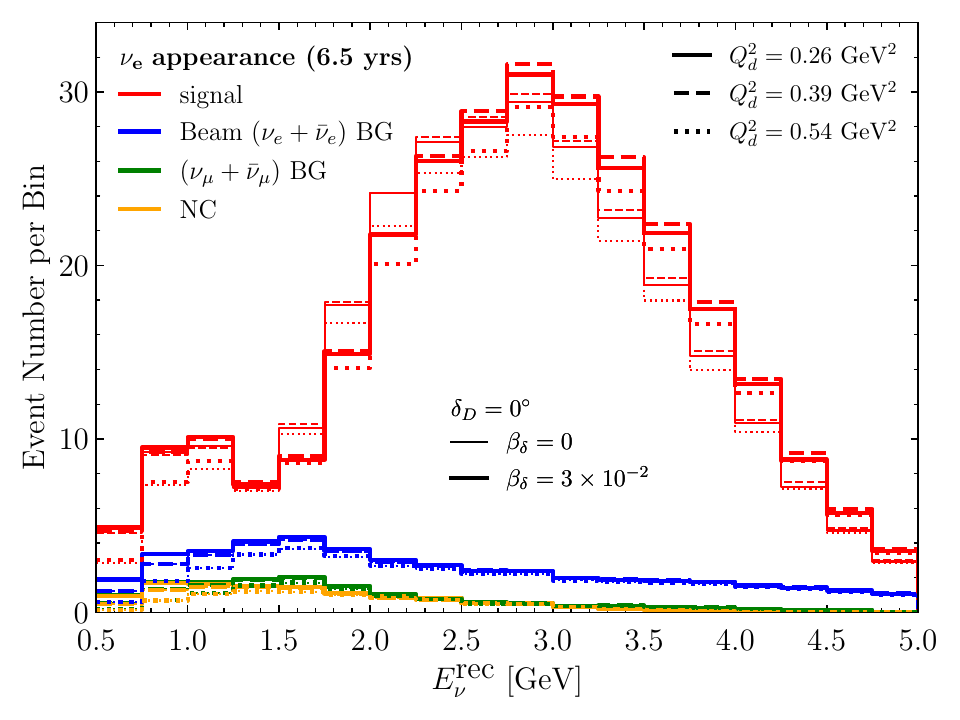}
\includegraphics[width=0.49\textwidth]{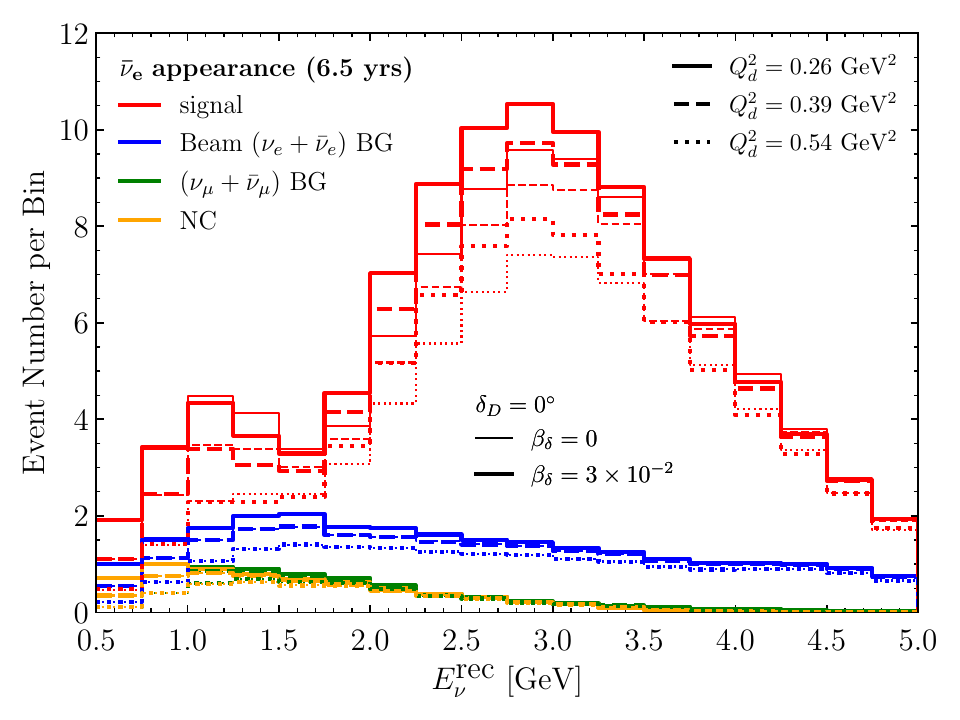}\\
\includegraphics[width=0.49\textwidth]{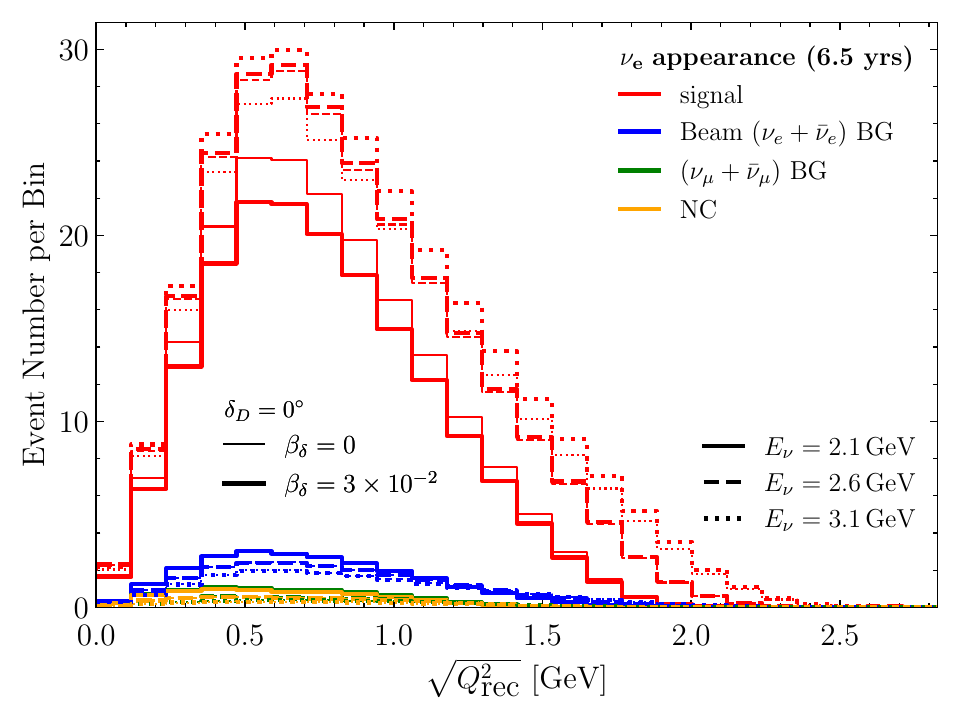}
\includegraphics[width=0.49\textwidth]{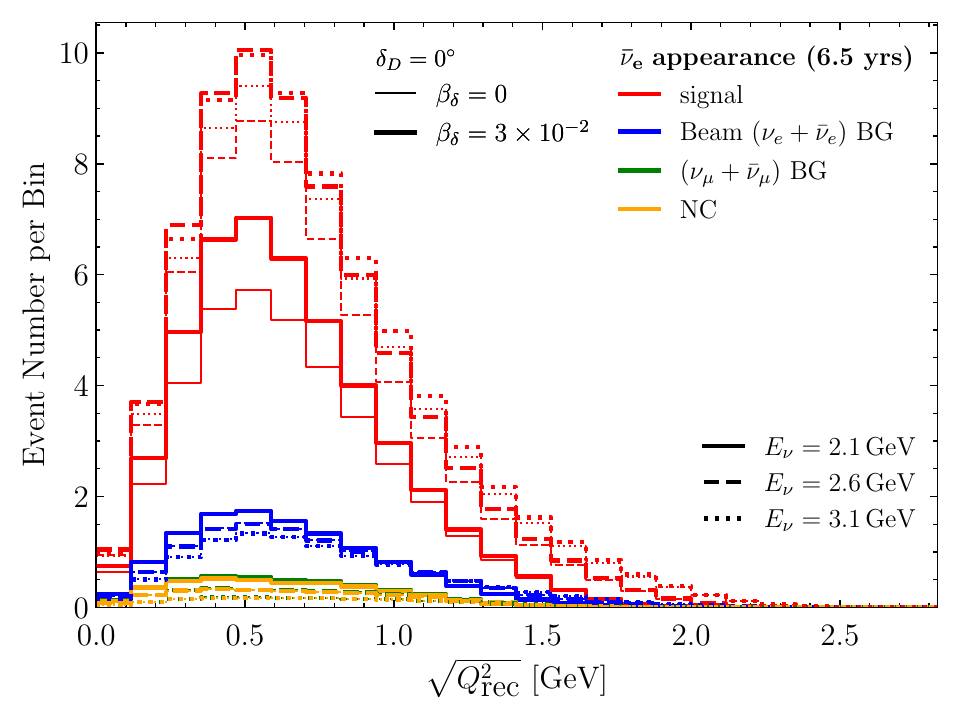}
\caption{
\textbf{Upper Panels:}
The reconstructed appearance event rates 
as a function of the reconstructed neutrino energy
$E_\nu^{\rm rec}$ with
detection momentum transfer $Q^2_d=0.26\,$GeV$^2$ (solid), 
$0.39\,$GeV$^2$ (dashed), and $0.54\,$GeV$^2$
(dotted) for the neutrino $\nu_e$ (upper left) and
anti-neutrino $\bar \nu_e$ (upper right) modes. 
\textbf{Lower Panels:} The reconstructed
appearance event rates 
as a function of $\sqrt{Q^2_{\rm rec}}$
with $E_\nu = 2.125\,$GeV (solid), $2.625\,$GeV
(dashed), and $3.125\,$GeV (dotted) for the
neutrino $\nu_e$ (lower left) and
anti-neutrino $\bar \nu_e$ (lower right) modes.
For all panels, the event rates are 
divided into four components: 
the signal (red), the intrinsic 
$\nu_e+\bar\nu_e$ beam background (blue), the
$\nu_\mu+\bar\nu_\mu$ beam background (green),
and the neutral current background (orange). 
We take $\delta_D=0^\circ$ and 
a 6.5 years of running time for each neutrino and
anti-neutrino mode to make illustration.}
\label{fig:Events}
\end{figure}
\subsection{Event Rate and CP Phase Sensitivity at DUNE}
\label{sec:erandCP}

\gfig{fig:Events} shows 
the event
rates at DUNE as function of the reconstructed
neutrino energy $E_\nu^{\rm rec}$ (upper panels) 
or momentum transfer $\sqrt{Q^2_{\rm rec}}$ (lower panels) for both 
neutrino (left panels) and 
anti-neutrino (right panels) modes. 
The event rate difference between the thin and
thick lines induced by 
the RG running effect can reach percentage level.
This happens for not just the signal (red lines)
but the various background event rates since all
channels are subject to the oscillation probabilities.
The most prominent channel is the beam $\nu_e + \bar \nu_e$
background that is modulated by $P_{ee}$. 
As pointed
out earlier at the end of \gsec{sec:vac-zd}, $P_{ee}$
is no longer independent of the Dirac CP phase
$\delta_D$ as
usually expected but becomes a function of the CP
phase difference $\Delta \delta_D$ induced by the
RG running effect. The largest variation
occurs in the ranges 
$[1.5, 4.5]$/$[1.25, 3.5]\,\rm{GeV}$ 
of $E_\nu^{\rm rec}$ 
and $[0.2, 1.5]$/$[0.2, 1.2]\,\rm{GeV}$ of
$\sqrt{Q^2_{\rm rec}}$ 
for the neutrino/anti-neutrino mode, respectively,
around the peaks.

\begin{figure}[t!]
\centering
\includegraphics[width=0.49\textwidth]{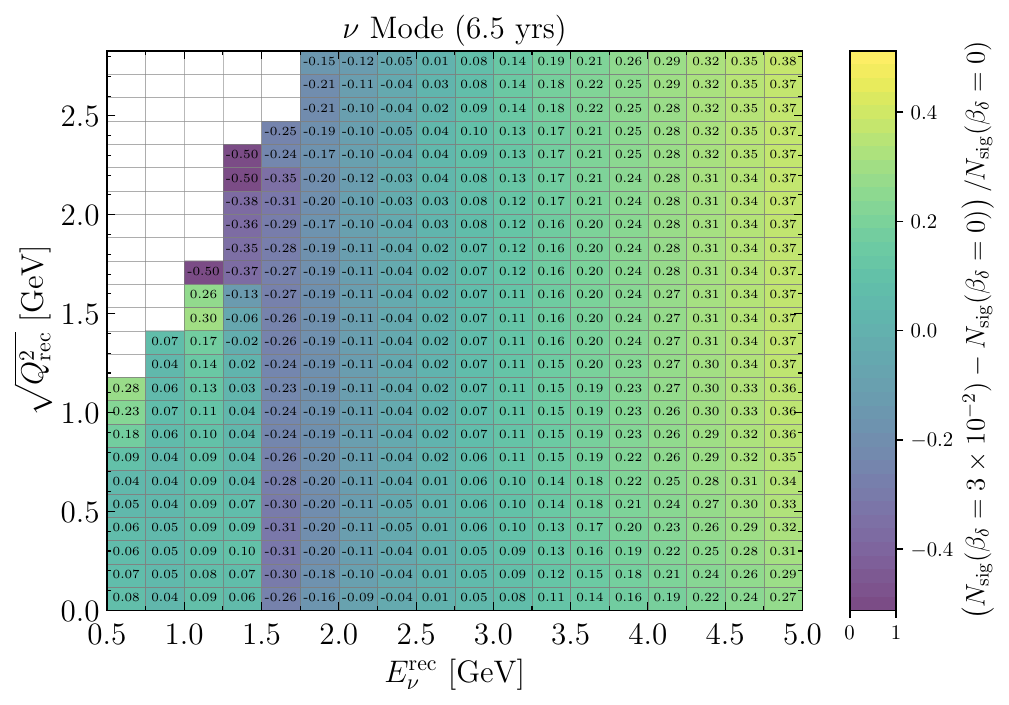}
\includegraphics[width=0.49\textwidth]{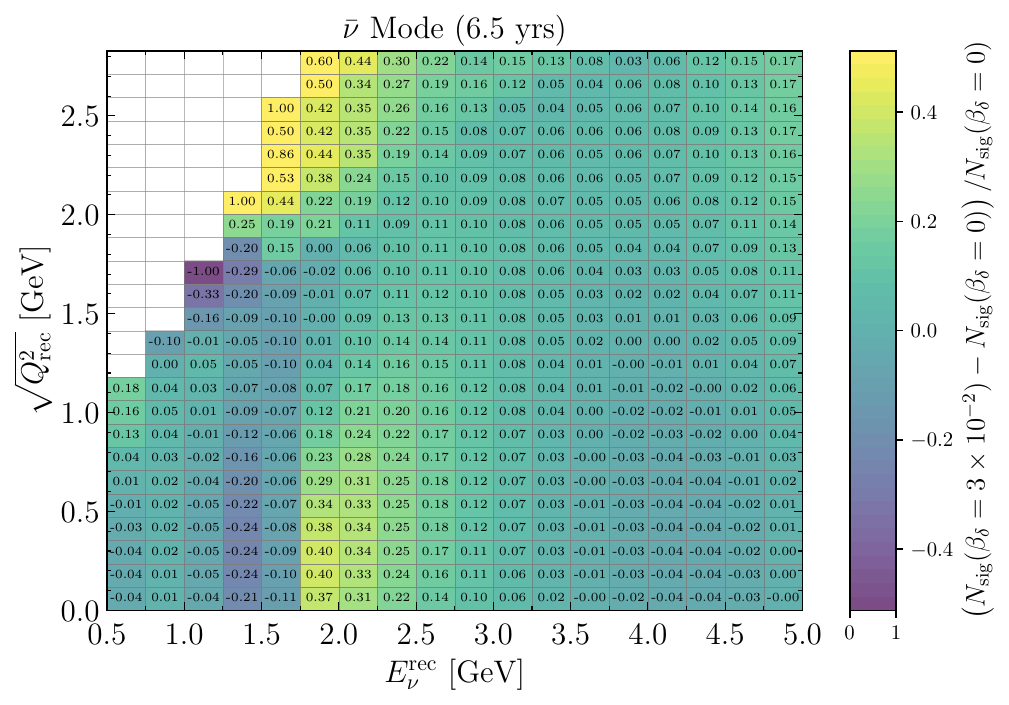}
\caption{ The relative difference,
$\Delta N/N \equiv \left(N_{\rm 
sig}(\beta_\delta= 3\times 10^{-2}) - N_{\rm sig}
(\beta_\delta=0)\right)/N_{\rm sig}
(\beta_\delta=0)$,
between the two cases with and without
the RG running effect in the 
two-dimensional $E_\nu^{\rm rec}$ 
and $\sqrt{Q^2_{\rm rec}}$ bins
for the neutrino (left) and
anti-neutrino (right) modes. 
We take $\delta_D=0^\circ$ and 
a 6.5 years of running time for each mode.}
\label{fig:2D-Events}
\end{figure}

To further explore the
features induced by the 
RG running effect, we show the relative
difference of the two-dimensional signal event rate,  
$\Delta N /N \equiv \left(N_{\rm 
sig} \left(\beta_\delta= 3\times 10^{-2}\right) 
- N_{\rm sig}
(\beta_\delta=0)\right)/$ $N_{\rm sig}
(\beta_\delta=0)$, on 
the $(E_\nu^{\rm
rec}, \sqrt{Q^2_{\rm rec}})$
plane in \gfig{fig:2D-Events}. 
Note that the 
%location of 
%peaks/depths 
distribution
of $\Delta N/N$ 
is quite different between 
the neutrino (left) and 
anti-neutrino (right) modes. 
For the neutrino mode, the positive 
$\Delta N/N$ values distribute
at $E_\nu^{\rm rec}\in [3.5,5]\,$GeV while
the negative one within 
$E\in [1.5, 2.5]\,$GeV.
As for the anti-neutrino mode,
the negative $\Delta N/N$ values mainly
distribute at $E_\nu^{\rm rec}\in [1,1.75]\,$GeV 
for most $\sqrt{Q^2_{\rm rec}}$ and $E_\nu^{\rm rec}\in [3.5, 4.75]\,$GeV
for $\sqrt{Q^2_{\rm rec}}\in [0,1.2]\,$GeV, with the positive values in the remaining part.
The RG running effect can 
be tested at the future LBL experiments 
such as DUNE. Unfortunately,
since the genuine CP phase is
undetermined, a degeneracy 
between the genuine CP phase $\delta_D$
and the RG running parameter 
$\beta_\delta$ arises which reduces
the CP phase sensitivity.
The structure in the two-dimensional
$(E_\nu^{\rm rec}, \sqrt{Q^2_{\rm rec}})$
distributions may help to disentangle the
degeneracy.

%The 2-D event rate provides 
%new detailed information in the search for the RG 
%running effect at DUNE experiment.

To quantify the impact on
the CP phase sensitivity from
the RG running effect at DUNE, 
we extend the $\chi^2$ function,
\begin{eqnarray}
  \chi^2 
=
  \chi^2_{\rm stat}+\chi^2_{\rm sys}+\chi^2_{\rm prior},
\label{eq:chi2-definition}
\end{eqnarray}
by incorporating multiple \texttt{glb} profiles.
More concretely, the first term $\chi^2_{\rm stat}$
for statistical fluctuations,
\begin{eqnarray}
      \chi^2_{\rm stat} 
   =
      \sum_{i'j'r} 
      \left(\frac{N_{i'j'}^r({\rm true}) - (1+ a_r) N_{i'j'}^{r, {\rm sig}}({\rm test})
    - (1+ b_r) N_{i'j'}^{r, {\rm bkg}}({\rm test})
      }{\sqrt{N_{i'j'}^r({\rm true})}} \right)^2 ,
      \quad 
\label{eq:chi2}
\end{eqnarray}
needs to sum over not just the
two indices $i'$ and $j'$ for the neutrino energy
and the momentum transfer but also
the third index $r$ for the various rules
from each experimental profile.
The quantity 
$N_{i'j'}^{r,{\rm sig}}$ ($N_{i'j'}^{r,{\rm bkg}}$) 
is the signal (background) event number within 
the $i'$-th reconstructed
energy as well as the $j'$-th 
reconstructed
momentum transfer bin
for the rule $r$ and can be 
calculated with 
true or test oscillation 
parameters. We define 
$N_{i'j'}^r({\rm true}) \equiv N_{i'j'}^{r,{\rm sig}}({\rm true}) +  
N_{i'j'}^{r,{\rm bkg}}({\rm true})$ 
to represent the total 
event numbers of signal
and background with the 
true oscillation 
parameters. Moreover, $a_r$ 
and $b_r$ are the signal 
and background normalization nuisance 
parameters, respectively, for the rule $r$.

The $\chi^2_{\rm sys}$ term 
contains the Gaussian 
priors of signal and 
background normalizations.
For the signal 
normalization, we take  
$\sigma_{\nu_e, \bar{\nu}_e}=2\%$ and 
$\sigma_{\nu_\mu, \bar{\nu}_\mu}=5\%$ 
for the electron (anti-)neutrino and muon (anti-)neutrino 
detection, respectively. A 
common $5\%$ uncertainty is 
used for the background 
normalization. These 
parameters are taken from 
the experiment simulation 
configurations of the
DUNE Technical Design 
Report \cite{Abi:2021arg}. 

The final term $\chi^2_{\rm prior}$ summarizes
the prior knowledge on the oscillation parameters.
We take their best-fit values from the global fit 
result \cite{deSalas:2020pgw},
\begin{subequations}
\begin{eqnarray}
&&
  \sin^2 \theta_s = 0.318,
\qquad
  \sin^2 \theta_a = 0.574,
\qquad
  \sin^2 \theta_r = 0.022,
\\
&&
  \Delta m^2_s = 7.50 \times 10^{-5}\,\mbox{eV}^2,
\qquad
  \Delta m^2_a = 2.55\times 10^{-3}\,\mbox{eV}^2.
\end{eqnarray}
\end{subequations}
Among these oscillation parameters, the solar 
mass squared difference $\Delta m_s^2$ and the solar 
mixing angle $\theta_s$ are fixed in our fit since
the oscillation probability variation induced by 
these two parameters is negligibly small 
\cite{Ge:2022iac}. This feature still applies even
in the presence of the RG running effect as we have
checked numerically. The reactor 
mixing angle $\theta_r$, the atmospheric mixing angle 
$\theta_a$, and the atmospheric mass-squared 
difference $\Delta m^2_a$ are treated as free 
parameters with priors taken 
from the marginalized 
one-dimensional $\chi^2$ curves \cite{deSalas:2020pgw}.
Due to large uncertainties and 
the existing tension between the T2K and 
NO$\nu$A results as mentioned in 
\gsec{sec:intro}, we do not include any 
prior on the Dirac CP phase $\delta_D$ and simply
treat it as a free parameter. Moreover, we take the 
normal ordering of neutrino masses
throughout our study.
We first consider a free $\beta_\delta$
while the prior knowledge 
can be extracted from the SBL experiments as we 
study in \gsec{sec:SBLconstraint}.

\begin{figure}[t!]
\centering
\includegraphics[width=0.48\textwidth]{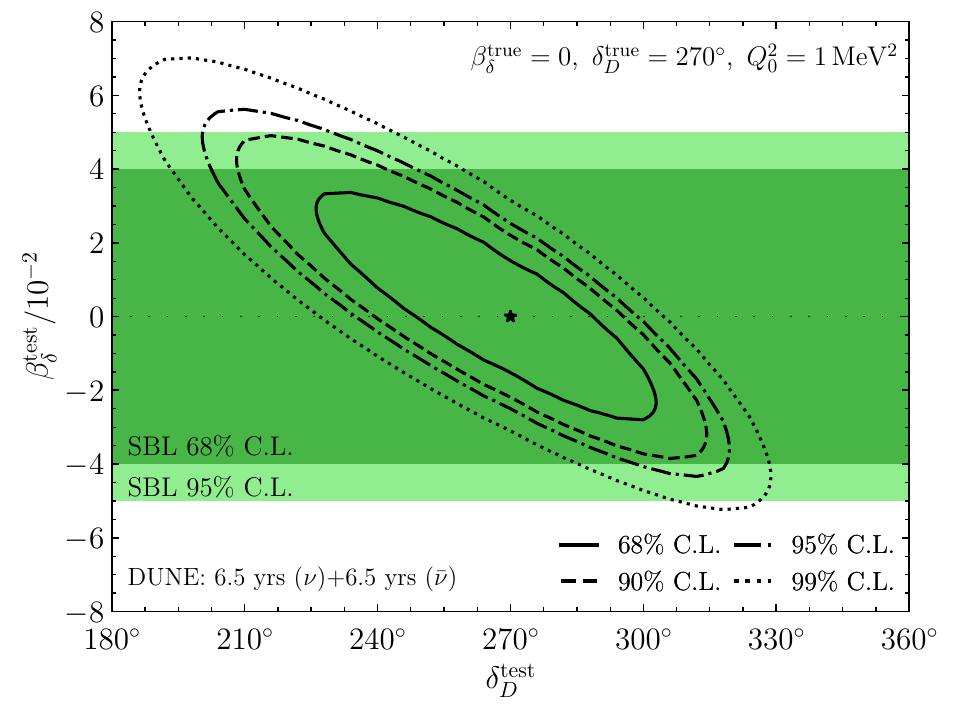}
\includegraphics[width=0.48\textwidth]{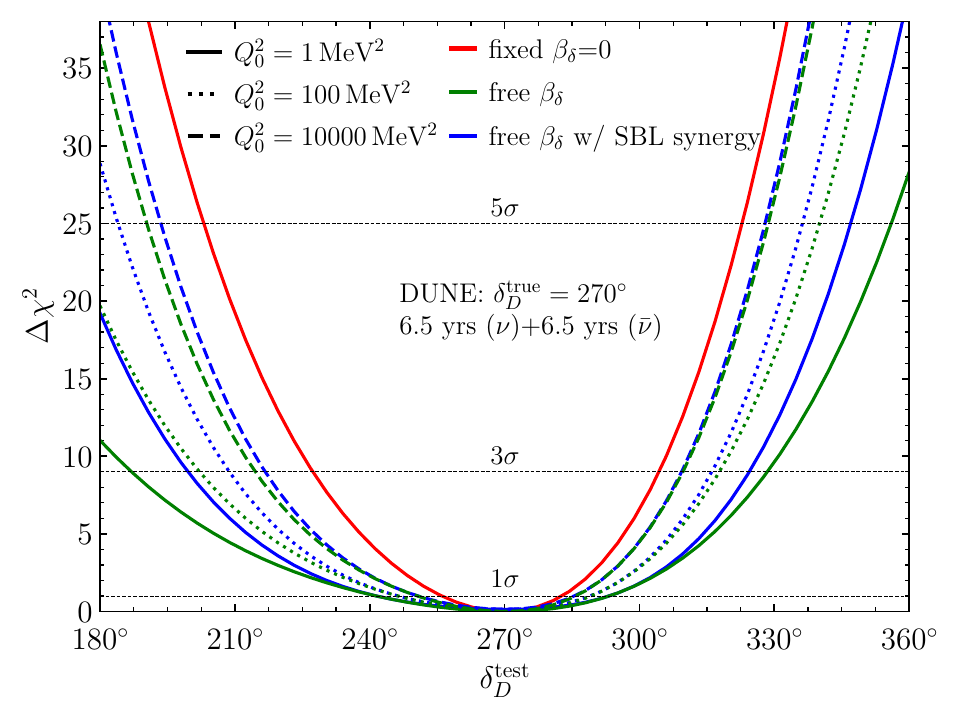}
\caption{
\textbf{Left:} The two-dimensional sensitivities on
$\delta_D$ and $\beta_\delta$ at the confidence
levels of $68\%$ (solid), $90\%$ (dashed), $95\%$ 
(dot-dashed), and $99\%$ (dotted) with $Q^2_0=1\,$MeV$^2$
and a 6.5 years running time each for the neutrino and
anti-neutrino modes. The true parameter setup 
$\delta^{\rm true}_D=270^\circ$ and 
$\beta^{\rm true}_\delta=0$ is marked by
a black star. 
Moreover, the SBL 
$\beta_\delta$ constraint at 
68\% (95\%) confidence 
level is shown by the 
dark-green (light-green) bands.
\textbf{Right:} The 
marginalized sensitivity to the
genuine CP phase $\delta_D$ with
three different prior cases: the fixed 
$\beta^{\rm test}_\delta=0$ 
(red), a totally free $\beta^{\rm test}_\delta$
(green), and a free 
$\beta^{\rm test}_\delta$ but 
with the SBL constraint (blue). 
For comparison, results
with $Q^2_0=1/100/10000\,$MeV$^2$
are shown as solid/dotted/dashed curves.
}
\label{fig:CP-sensitivity}
\end{figure}

We perform the $\chi^2$ analysis according to
\geqn{eq:chi2-definition} to calculate 
the DUNE sensitivities to 
$\delta_D$ and $\beta_\delta$ as shown 
in the left panel of \gfig{fig:CP-sensitivity}. 
Based on the true oscillation 
parameter setup
($\delta_D=270^\circ$ and 
$\beta_\delta=0$) that is 
marked by a black star,
the black contours show the sensitivities at
confidence levels of
$68\%$ (solid), $90\%$
(dashed), $95\%$ 
(dot-dashed), and 
$99\%$ (dotted).
Those points on the same contour have exactly
the same $\Delta \chi^2$ value and hence degenerate
with each other.
Since a positive (negative) 
$\beta_\delta$ can
increase (decrease)
$\delta_{\rm D}$ according to
\geqn{eq:new_parametrization}, the 
combination of a positive 
$\beta_\delta$ and a 
smaller $\delta_D$
degenerates with a
combination of opposite values. 
This degeneracy correlates 
the two parameters
and makes the sensitivity 
contours tilt from top left (larger $\beta_\delta$ and smaller $\delta_D$)
to bottom right 
(smaller $\beta_\delta$ and larger $\delta_D$). 
The tilting behavior 
implies that the CP phase sensitivity
would be reduced from the case without RG running. 
Moreover, there is an asymmetry in the sensitivity 
contour around the black star. 
The asymmetry is not caused by
the RG running effect since it already
exists between the intersection points of the
contours and the horizontal line for
$\beta^{\rm test}_\delta = 0$.
In fact, it arises from the matter effect 
while the RG running case inherits and
enlarges this asymmetry which can also
been seen from the right panel.

To clearly see the impact of RG running
on the CP phase sensitivity,
the marginalized $\Delta \chi^2$ is shown 
in the right panel of 
\gfig{fig:CP-sensitivity}.
The red curve with a fixed
$\beta_\delta^{\rm test}=0$
corresponds to the 
standard oscillation
case without RG running while the green one
sets $\beta_\delta^{\rm test}$ free.
From the red to the green curves, the CP phase sensitivity 
significantly reduces. 
The original $1\sigma$ interval of 
$\sim[255^\circ, 283^\circ]$ 
changes to $\sim[240^\circ, 295^\circ]$
with almost a factor of 2 reduction.

Moreover, we investigate the effect of the reference
momentum transfer $Q^2_0$ by 
considering three different values: 
$Q^2_0=1\,$MeV$^2$ (solid), 
100\,MeV$^2$ (dotted), 
and 10000\,MeV$^2$ (dashed). 
Although the CP phase sensitivity
increases with $Q^2_0$ since the
distance from the actual momentum transfer $Q^2_d$
decreases to reduce the RG running effect,
the impact on the CP measurement at DUNE
still cannot be neglected
even for $Q^2_0=10000\,$MeV$^2$.
Therefore, there exists a 
large parameter space for the 
energy scale of new physics 
that can significantly 
affect the leptonic CP phase 
sensitivity.

\section{Synergy with Existing Short-Baseline Experiments}
\label{sec:SBLconstraint}

In the zero-distance limit,
the probability of flavor transition
induced by the RG running effect
depends on the  
CP phase difference $\Delta\delta_D$
instead of the genuine CP phase $\delta_D$
as shown in \geqn{eq:sblEQ}. 
Hence SBL experiments can
provide a clean measurement for RG running.
Therefore, a synergy between the long- and 
short-baseline experiments can
disentangle the genuine
CP phase ($\delta_D$) from the RG 
running effect ($\beta_\delta$).

The searches for neutrino transition
in the zero-distance limit
are done by $\mathcal{O}(100\,\rm{m})$ SBL 
experiments such 
as ICARUS \cite{ICARUS:2013cwr}, CHARM-II \cite{CHARMII:1994rnc}, NOMAD \cite{NOMAD:2003mqg}, and NuTeV \cite{NuTeV:2002daf}.
All these four experiments use both
$\nu_\mu$ and $\nu_e$ neutrinos
as well as the anti-neutrino counterparts.
Since ICARUS has only 4 detected $\nu_e$ events from 
the result \cite{ICARUS:2013cwr},
we will not take it into account due to lack 
of statistical significance.
As for the other three
(CHARM-II, NOMAD, and NuTeV),
$\mathcal{O}(1000)$ $\nu_e$ events
are detected at each experiment
\cite{CHARMII:1994rnc,NOMAD:2003mqg,NuTeV:2002daf}.
Note that anti-neutrinos 
are also detected at CHARM-II and NuTeV.
Since the anti-neutrino case is similar to
the neutrino one, we just discuss 
the neutrino case at SBL
experiments for illustration in the 
following text. These
$\nu_e$ neutrinos are from two sources: 
the $\nu_e$ disappearance from the $\nu_e$
flux and those from the 
$\nu_\mu\rightarrow\nu_e$ transition
due to the zero-distance effect.
The experimental constraint on the 
$\nu_\mu\rightarrow\nu_e$
transition probability 
can be obtained by comparing the predicted $\nu_e$ neutrino 
spectra with the experimental data.
Besides the $\nu_\mu\rightarrow\nu_e$ transition, 
these SBL experiments can also search for
$\nu_e\rightarrow\nu_\tau$ as well as 
$\nu_\mu\rightarrow\nu_\tau$ \cite{CHARM-II:1993nip,NOMAD:2001xxt,CCFRNuTeV:1998gjj}.
However, since the charged $\tau$ lepton produced 
from $\nu_\tau$ CC scattering
decays immediately into final-state particles 
such as $e$, $\mu$, $\pi$, and $K$ that
can also be produced in other CC or
NC events, the detection of $\nu_\tau$
suffers from these potential backgrounds.
Moreover, the $\tau$ selection criteria
include much more kinematic requirements
than its counterparts of $e$,
which needs more complicated analysis. 
The SBL experiments find no evidence of
$\nu_\tau$ with large background uncertainties.
Hence we do not consider 
such $\nu_{e,\mu}\rightarrow\nu_\tau$ 
transition channels in our current 
study.

To obtain the experimental constraint
on $\beta_\delta$,
we first calculate 
the predicted $\nu_e$ event rate
with the RG running effect.
For a neutrino experiment,
the predicted $\nu_e$ event rate 
can be described as
$N_{\nu_e} = A \sum_{\alpha}\phi(\nu_\alpha)P_{\alpha e}\sigma(\nu_e)$
with $A$ being the coefficient that
is related to the experimental 
efficiencies and reconstruction errors of 
the detector, $\phi(\nu_\alpha)$ the 
$\alpha$ flavor neutrino flux, 
$P_{\alpha e}$ the
$\nu_\alpha\rightarrow\nu_e$ oscillation
probability, and $\sigma(\nu_e)$ the
cross section for scattering 
between $\nu_e$ and target. The predicted
$\nu_e$ event rate $R(\nu_e)$ from the collaboration 
\cite{CHARMII:1994rnc,NOMAD:2003mqg,NuTeV:2002daf}
was obtained by assuming no oscillation,
$P_{\alpha e}=\delta_{\alpha e}$.
As for the case with RG running, 
we calculate
the predicted $\nu_e$ event rates 
$R'(\nu_e)$ by
multiplying the transition 
probability in \geqn{eq:sblEQ} 
with the $\nu_e$ and 
$\nu_\mu$ MC neutrino
event rates ($R(\nu_e)$ and 
$R(\nu_\mu)$),
$R'(\nu_e)\equiv P_{\nu_e \rightarrow \nu_e}  R(\nu_e)+P_{\nu_\mu \rightarrow \nu_e}  R(\nu_\mu)$.
We take $\sigma(\nu_e)\approx \sigma(\nu_\mu)$ 
since the neutrino energy 
is much larger than the charged lepton masses. 

Note that the transition probabilities $P_{\nu_e \rightarrow \nu_e} $ and 
$P_{\nu_\mu \rightarrow \nu_e} $ with 
the RG running effect are
$Q^2_d$ dependent as discussed in \gsec{sec:Q2dependentP}.
So the typical high-energy neutrinos at SBL experiments
have the advantage of producing large
momentum transfer to enhance the RG running effect.
However, there
is no such momentum transfer information in the published
SBL experimental data 
\cite{CHARMII:1994rnc,NOMAD:2003mqg,NuTeV:2002daf}.
So in the  
transition probability calculation, 
we take the averaged oscillation probability 
by integrating \geqn{eq:sblEQ} over the 
$Q^2_d$ distribution extracted from the 
GENIE simulation.

\begin{figure}[t!]
\centering
\includegraphics[scale = 0.56]{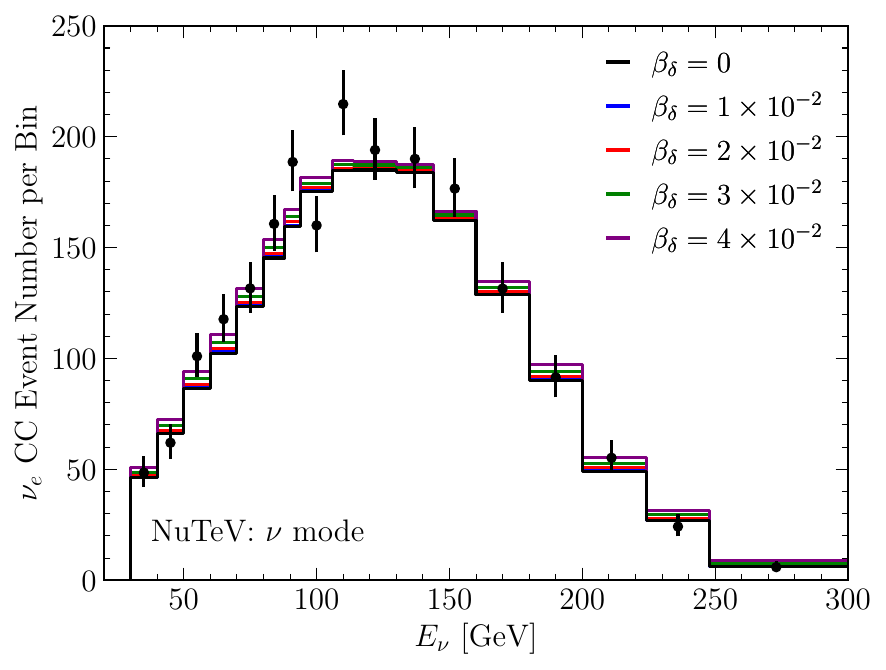}
\includegraphics[scale = 0.56]{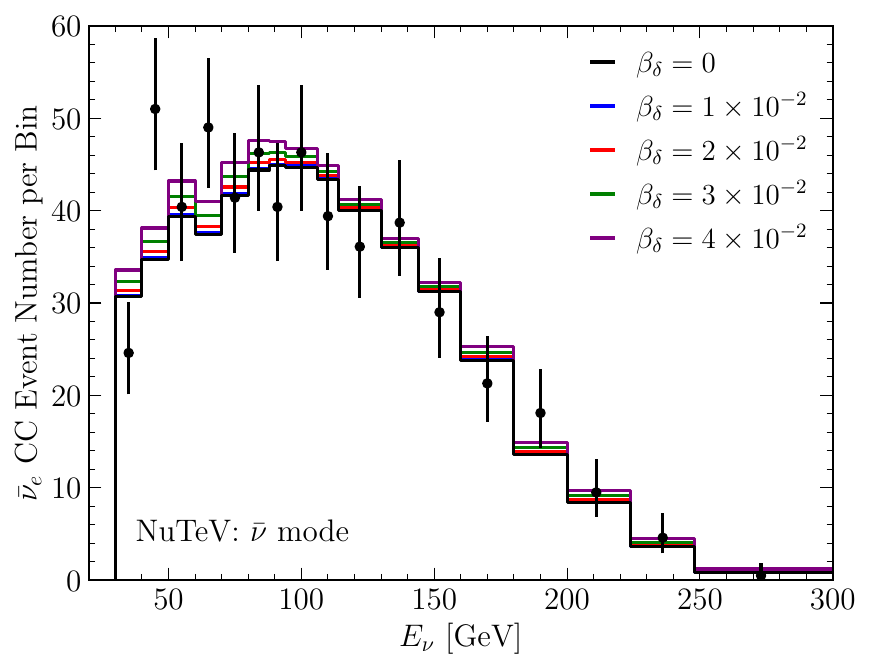}\\
\includegraphics[scale = 0.56]{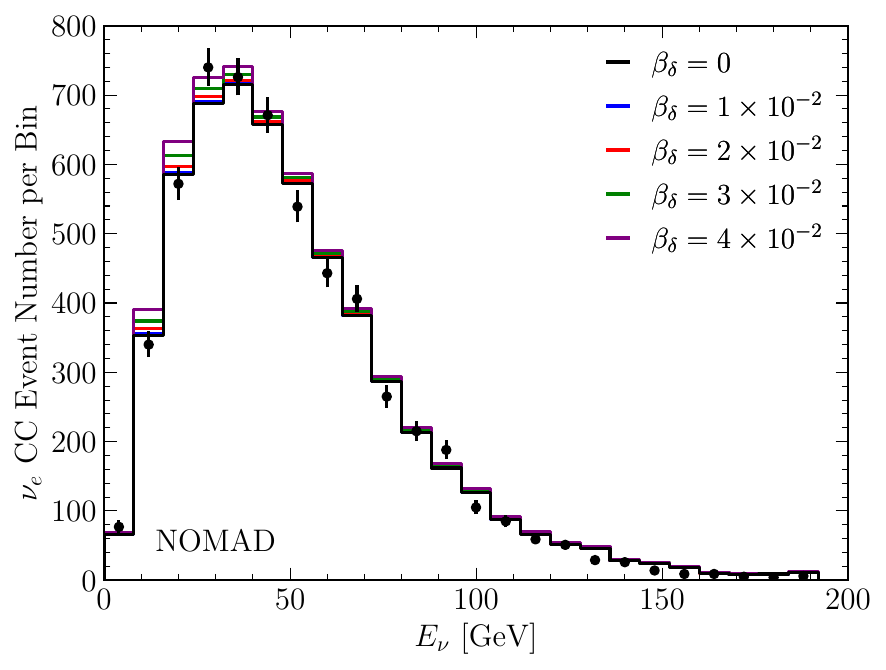}
\includegraphics[scale = 0.56]{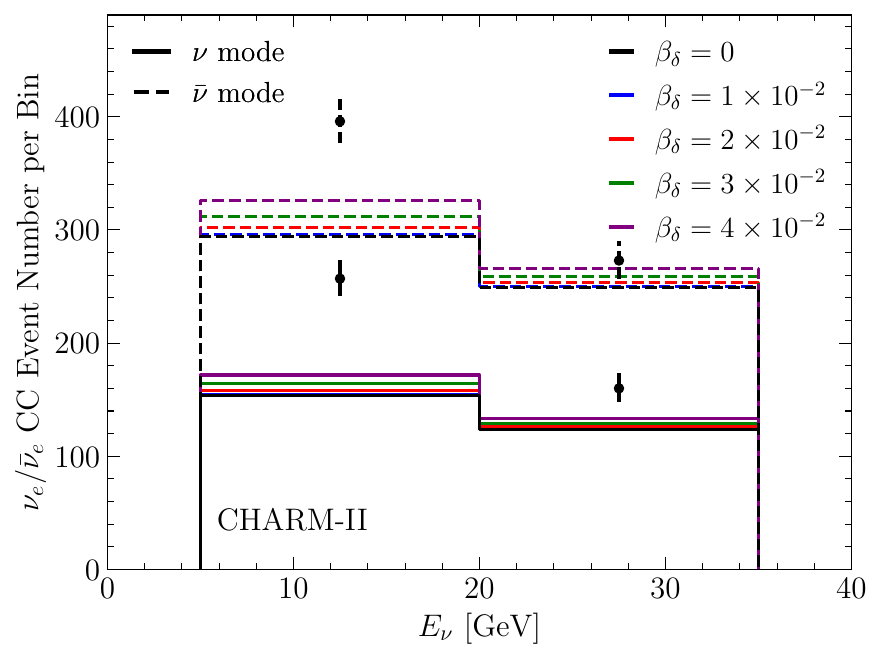}
\caption{
The $\nu_e / \bar \nu_e$
CC event rates for NuTeV with the neutrino mode
(upper left), 
NuTeV with the anti-neutrino mode (upper 
right), NOMAD with the 
neutrino mode (lower left), 
and CHARM-II with both modes (lower right), respectively. 
Besides the data points (black dot with error bar),
the RG running effect is shown with
$\beta_\delta=0$ (black), $1\times 10^{-2}$
(blue), $2\times 10^{-2}$ (red), $3\times 
10^{-2}$ (green), and $4\times 10^{-2}$
(purple). 
}
\label{fig:SBL-events}
\end{figure}

\gfig{fig:SBL-events} shows the 
predicted electron-flavor neutrino 
event rates as well as the 
experimental data from NuTeV, NOMAD, 
and CHARM-II. 
For all experiments, the predicted 
event rates with a percentage level $\beta_\delta$   
are roughly $\mathcal{O} (10\%)$ more 
than the MC prediction without 
flavor transition. 
Note that flipping the sign of 
$\beta_\delta$ does not
affect the predicted event rates
since $\beta_\delta$ only appears 
as the squared term $\sin^2(\Delta \delta_D/2)$ in \geqn{eq:sblEQ}.
Such variations 
are comparable with the experimental 
uncertainties which means 
the experimental sensitivity to
$\beta_\delta$ can reach 
percentage level.
Moreover, with the predicted event 
rates being a linear combination of both  
the $\nu_e$ MC neutrino event 
rate $R(\nu_e)$ and the $\nu_\mu$ 
counterpart $R(\nu_\mu)$, 
the largest variation
between the predicted and expected 
event rates does not exactly 
occur at the $\nu_e$
event rate peak position since the
$\nu_e$ and $\nu_\mu$ fluxes peak at
different energies
\cite{CHARMII:1994rnc,NOMAD:2003mqg,NuTeV:2002daf}.

\begin{figure}[t!]
\centering
\includegraphics[scale = 0.8]{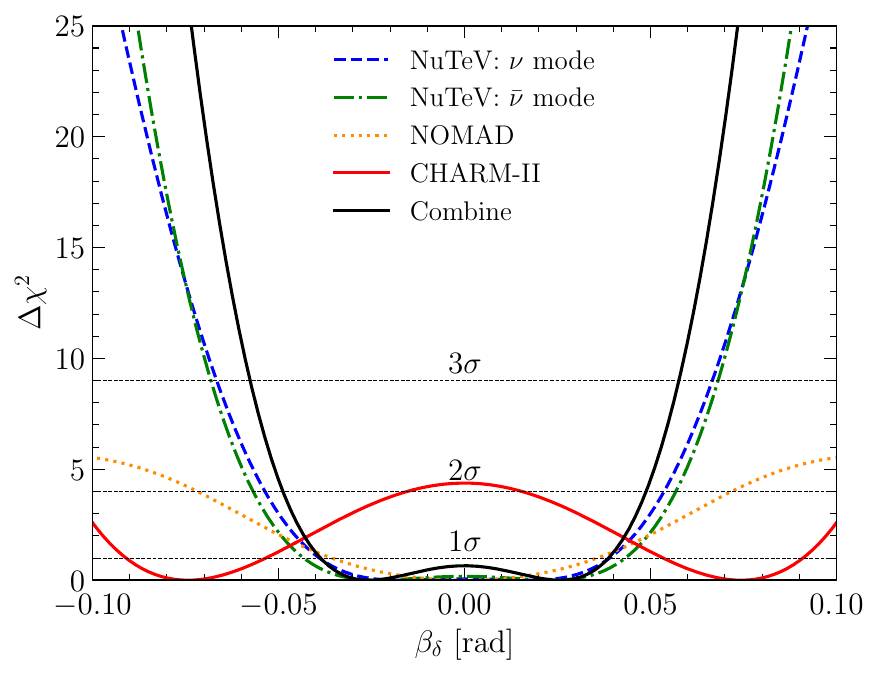}
\caption{
The $\Delta\chi^2$ ($\equiv \chi^2-\chi^2_{\rm min}$) as
a function 
of $\beta_\delta$ for NuTeV with the 
neutrino mode (dashed blue), 
NuTeV with the anti-neutrino mode (dot-dashed green), NOMAD with the
neutrino mode (dotted orange),
CHARM-II with both neutrino and anti-neutrino modes (solid red), 
and their combination (solid black). 
Moreover, the $1\sigma$, $2\sigma$, and 
$3\sigma$ confidence levels are 
shown by horizontal black dotted lines.
}
\label{fig:SBL}
\end{figure}

To obtain the sensitivity on $\beta_\delta$, 
we consider the following $\chi^2$ function,
\begin{align}
  \chi^2
\equiv &
  \sum_{i = 1}^{n_{\rm bins}} 
  \left(
    \frac{
        N^{{\rm data},i}
      -
        (1+ b^e)
        N^{{\rm fit},i}_{\nu_e\rightarrow \nu_e}
      -(1+ b^\mu)
        N^{{\rm fit},i}_{\nu_\mu\rightarrow \nu_e}
      }{\sqrt{N^{{\rm data},i} + (N^{{\rm fit},i}_{\nu_e \rightarrow \nu_e} \sigma_i^e)^2 + (N^{{\rm fit},i}_{\nu_\mu \rightarrow \nu_e} \sigma_i^\mu)^2}}
  \right)^2 
+ \left(\frac{b^e}{\sigma_e}\right)^2
+ \left(\frac{b^\mu}{\sigma_\mu}\right)^2,
  \label{eq:chi2_def}
\end{align}
where $N^{{\rm data},i}$ and $N^{{\rm fit},i}_{\nu_\alpha \rightarrow \nu_e}$ are the 
experimental data and predicted event 
number in the $i$-th energy bin, respectively.
The nuisance parameter $b^e$ ($b^{\mu}$)
describes the total 
normalization uncertainty $\sigma_e$ ($\sigma_\mu$) for the 
electron-flavor (muon-flavor) neutrino 
flux. Moreover,
the parameters $\sigma_i^{e}$ 
and $\sigma_i^{\mu}$ are the 
systematic uncertainties in each energy 
bin 
\cite{CHARMII:1994rnc,NOMAD:2003owt,
NuTeV:2002daf}. We obtain the constraint 
on $\beta_\delta$ 
by minimizing $\chi^2$ 
over the nuisance parameters and subtracting the 
global minimum value $\chi^2_{\rm min}$,
$\Delta \chi^2\equiv \chi^2-\chi^2_{\rm min}$,
as shown in \gfig{fig:SBL}.  
For NuTeV with both $\nu$ (dashed blue) and $\bar\nu$ 
modes (dot-dashed green), the $\Delta\chi^2$ curve 
is flat in the region of 
$|\beta_\delta|\lesssim 4\times 10^{-2}$, 
but raises quickly at larger values. The NOMAD 
experiment (dotted orange) is more sensitive to smaller
$|\beta_\delta|$ than NuTeV with higher 
statistics. For NOMAD and the $\nu$ mode at NuTeV,
the $\Delta\chi^2$ minimum lies at 
$\beta_\delta = 0$, while the $\bar\nu$ mode 
at NuTeV slightly 
prefers a non-zero $\beta_\delta$. 
Moreover, the CHARM-II 
experiment (solid red) disfavors $\beta_\delta=0$
by $\sim 2\sigma$ since the observed 
data points have an excess over the
expected event rates as shown in 
\gfig{fig:SBL-events}. 
Note that 
all these $\Delta \chi^2$ curves
are symmetric around $\beta_\delta=0$ 
since the predicted event rates do not depend on
the sign of $\beta_\delta$ as explained before.
The combined result (solid black) of these
experiments gives a bound of
$|\beta_\delta| \lesssim 4\times10^{-2}$ 
($5\times 10^{-2}$) at $68\%$ ($95\%$) 
confidence level with the 
best fit being $|\beta_\delta|\approx 2.5\times 10^{-2}$.

For comparison, we show the 68\% and 95\% sensitivities
on $\beta_\delta$ as dark- and light-green bands in
the left panel of \gfig{fig:CP-sensitivity}. Since the
$\beta_\delta$ sensitivities at SBL experiments is
actually comparable with the one at DUNE, their
synergy should further improve the sensitivity.
As the SBL experiments (NuTeV, NOMAD, and CHARM-II)
already have existing data, their constraints on
$\beta_\delta$ can be treated as priors.
The $\Delta \chi^2$ curves shown in the right panel
of \gfig{fig:CP-sensitivity} does improve after taking the
SBL priors into consideration. Note that the current
priors at SBL experiments are obtained without considering
the momentum transfer distribution due to lack of
information. A more detailed analysis with full
data from the experimental side should significantly
enhance the existing priors.

\section{Conclusion \& Discussion}
\label{sec:conclusion}

We establish a general and complete formalism for
studying the RG running effect on the neutrino
oscillation. Especially, our formalism allows an
energy dependent momentum transfer for the forward
scattering and hence the matter potential. With a
broad momentum transfer distribution instead of a
single fixed value,
our study shows that the degeneracy between
the RG running parameter $\beta_\delta$ and the
genuine CP phase $\delta_D$ will significantly reduce
the CP phase sensitivity at DUNE. With mismatch between
the momentum transfers in the neutrino production
and detection processes, the transition probability
of the appearance channels does not vanish even in
the limit of zero distance. Consequently, the
short-baseline experiments can provide an independent
measurement of the RG running parameter $\beta_\delta$
and its synergy with the long-baseline DUNE can help
to disentangle the degeneracy and restore the CP phase
sensitivity to some extent. Due to lack of information
on the momentum transfer, the $\beta_\delta$
sensitivity at short-baseline experiments
(NuTeV, NOMAD, and CHARM-II) is comparable with
the one at DUNE. We expect more disentangling
capability with full data and detailed analysis
from the experimental side.

\section*{Acknowledgements}
The authors are supported by the National Natural Science
Foundation of China (12375101, 12090060 and 12090064) and the SJTU Double First
Class start-up fund (WF220442604).
SFG is also an affiliate member of Kavli IPMU, University of Tokyo.
 PSP is also supported by the Grant-in-Aid for Innovative Areas No. 19H05810.

\addcontentsline{toc}{section}{References}
\bibliographystyle{plain}

\begin{thebibliography}{99}

\bibitem{Canetti:2012zc}
L.~Canetti, M.~Drewes and M.~Shaposhnikov,
``{\it Matter and Antimatter in the Universe},''
\href{https://doi.org/10.1088/1367-2630/14/9/095012}{New J. Phys. \textbf{14}, 095012 (2012)}
[\href{https://arxiv.org/abs/1204.4186}{arXiv:1204.4186} [hep-ph]].

\bibitem{Sakharov:1967dj}
A.~D.~Sakharov,
``{\it Violation of CP Invariance, C asymmetry, and baryon asymmetry of the universe},''
\href{https://doi.org/10.1070/PU1991v034n05ABEH002497}{Pis'ma Zh. Eksp. Teor. Fiz. \textbf{5}, 32-35 (1967)}

\bibitem{Gavela:1993ts}
M.~B.~Gavela, P.~Hernandez, J.~Orloff and O.~Pene,
``{\it Standard model CP violation and baryon asymmetry},''
\href{https://doi.org/10.1142/S0217732394000629}{Mod. Phys. Lett. A \textbf{9}, 795-810 (1994)}
[\href{https://arxiv.org/abs/hep-ph/9312215}{arXiv:hep-ph/9312215} [hep-ph]].

\bibitem{Huet:1994jb}
P.~Huet and E.~Sather,
``{\it Electroweak baryogenesis and standard model CP violation},''
\href{https://doi.org/10.1103/PhysRevD.51.379}{Phys. Rev. D \textbf{51}, 379-394 (1995)}
[\href{https://arxiv.org/abs/hep-ph/9404302}{arXiv:hep-ph/9404302} [hep-ph]].

\bibitem{Gavela:1994dt}
M.~B.~Gavela, P.~Hernandez, J.~Orloff, O.~Pene and C.~Quimbay,
``{\it Standard model CP violation and baryon asymmetry. Part 2: Finite temperature},''
\href{https://doi.org/10.1016/0550-3213(94)00410-2}{Nucl. Phys. B \textbf{430}, 382-426 (1994)}
[\href{https://arxiv.org/abs/hep-ph/9406289}{arXiv:hep-ph/9406289} [hep-ph]].

\bibitem{Fukugita:1986hr}
  M.~Fukugita and T.~Yanagida,
  ``{\it Baryogenesis Without Grand Unification},''
  \href{http://dx.doi.org/10.1016/0370-2693(86)91126-3}
  {Phys. Lett. B \textbf{174}, 45-47 (1986)}.

\bibitem{Buchmuller:2005eh}
W.~Buchmuller, R.~D.~Peccei and T.~Yanagida,
``{\it Leptogenesis as the origin of matter},''
\href{https://doi.org/10.1146/annurev.nucl.55.090704.151558}{
Ann. Rev. Nucl. Part. Sci. \textbf{55}, 311-355 (2005)
}
[\href{https://arxiv.org/abs/hep-ph/0502169}{arXiv:hep-ph/0502169} [hep-ph]].

\bibitem{Davidson:2008bu}
S.~Davidson, E.~Nardi and Y.~Nir,
``{\it Leptogenesis},''
\href{https://doi.org/10.1016/j.physrep.2008.06.002}{
Phys. Rept. \textbf{466}, 105-177 (2008)
}
[\href{https://arxiv.org/abs/0802.2962}{arXiv:0802.2962} [hep-ph]].

\bibitem{PDG22-NuCP}
   M.~C.~Gonzalez-Garcia and M.~Yokoyam,
  ``{\it Neutrino Masses, Mixing, and Oscillations}'',
  (Chapter 14) of
  R.~L.~Workman \textit{et al.} [Particle Data Group],
  ``{\it Review of Particle Physics},''
  \href{https://doi.org/10.1093/ptep/ptac097}{PTEP \textbf{2022}, 083C01 (2022)}.

\bibitem{Pascoli:2006ie}
S.~Pascoli, S.~T.~Petcov and A.~Riotto,
``{\it Connecting low energy leptonic CP-violation to leptogenesis},''
\href{https://doi.org/10.1103/PhysRevD.75.083511}{Phys. Rev. D \textbf{75}, 083511 (2007)}
[\href{https://arxiv.org/abs/hep-ph/0609125}{arXiv:hep-ph/0609125} [hep-ph]].

\bibitem{Branco:2006ce}
G.~C.~Branco, R.~Gonzalez Felipe and F.~R.~Joaquim,
``{\it A New bridge between leptonic CP violation and leptogenesis},''
\href{https://doi.org/10.1016/j.physletb.2006.12.060}{Phys. Lett. B \textbf{645}, 432-436 (2007)}
[\href{https://arxiv.org/abs/hep-ph/0609297}{arXiv:hep-ph/0609297} [hep-ph]].

\bibitem{Pascoli:2006ci}
S.~Pascoli, S.~T.~Petcov and A.~Riotto,
``{\it Leptogenesis and Low Energy CP Violation in Neutrino Physics},''
\href{https://doi.org/10.1016/j.nuclphysb.2007.02.019}{Nucl. Phys. B \textbf{774}, 1-52 (2007)}
[\href{https://arxiv.org/abs/hep-ph/0611338}{arXiv:hep-ph/0611338} [hep-ph]].

\bibitem{Anisimov:2007mw}
A.~Anisimov, S.~Blanchet and P.~Di Bari,
``{\it Viability of Dirac phase leptogenesis},''
\href{https://doi.org/10.1088/1475-7516/2008/04/033}{JCAP \textbf{04}, 033 (2008)}
[\href{https://arxiv.org/abs/0707.3024}{arXiv:0707.3024} [hep-ph]].

\bibitem{Molinaro:2009lud}
E.~Molinaro and S.~T.~Petcov,
``{\it The Interplay Between the 'Low' and 'High' Energy CP-Violation in Leptogenesis},''
\href{https://doi.org/10.1140/epjc/s10052-009-0985-3}{Eur. Phys. J. C \textbf{61}, 93-109 (2009)}
[\href{https://arxiv.org/abs/0803.4120}{arXiv:0803.4120} [hep-ph]].

\bibitem{Ge:2010js}
  Shao-Feng Ge, Hong-Jian He and Fu-Rong Yin,
  ``{\it Common Origin of Soft mu-tau and CP Breaking in Neutrino Seesaw and the Origin of Matter},''
  \href{http://dx.doi.org/10.1088/1475-7516/2010/05/017}
  {JCAP \textbf{05}, 017 (2010)}
  [\href{http://arxiv.org/abs/1001.0940}{arXiv:1001.0940} [hep-ph]].

\bibitem{Dolan:2018qpy}
M.~J.~Dolan, T.~P.~Dutka and R.~R.~Volkas,
``{\it Dirac-Phase Thermal Leptogenesis in the extended Type-I Seesaw Model},''
\href{https://doi.org/10.1088/1475-7516/2018/06/012}{JCAP \textbf{06}, 012 (2018)}
[\href{https://arxiv.org/abs/1802.08373}{arXiv:1802.08373} [hep-ph]].

\bibitem{NOvA:2019cyt}
M.~A.~Acero \textit{et al.} [NOvA],
``{\it First Measurement of Neutrino Oscillation Parameters using Neutrinos and Antineutrinos by NOvA},''
\href{https://doi.org/10.1103/PhysRevLett.123.151803}{Phys. Rev. Lett. \textbf{123}, no.15, 151803 (2019)}
[\href{https://arxiv.org/abs/1906.04907}{arXiv:1906.04907} [hep-ex]].

\bibitem{T2K:2019bcf}
K.~Abe \textit{et al.} [T2K],
``{\it Constraint on the matter\textendash{}antimatter symmetry-violating phase in neutrino oscillations},''
\href{https://doi.org/10.1038/s41586-020-2177-0}{Nature \textbf{580}, no.7803, 339-344 (2020)}
[erratum: \href{https://doi.org/10.1038/s41586-020-2415-5}{Nature \textbf{583}, no.7814, E16 (2020)}]
[\href{https://arxiv.org/abs/1910.03887}{arXiv:1910.03887} [hep-ex]].

\bibitem{NOvA:2004blv}
D.~S.~Ayres \textit{et al.} [NOvA],
``{\it NOvA: Proposal to Build a 30 Kiloton Off-Axis Detector to Study $\nu_{\mu} \to \nu_e$ Oscillations in the NuMI Beamline},''
[\href{https://arxiv.org/abs/hep-ex/0503053}{arXiv:hep-ex/0503053} [hep-ex]].

\bibitem{T2K:2011qtm}
K.~Abe \textit{et al.} [T2K],
``{\it The T2K Experiment},''
\href{https://doi.org/10.1016/j.nima.2011.06.067}{Nucl. Instrum. Meth. A \textbf{659}, 106-135 (2011)}
[\href{https://arxiv.org/abs/1106.1238}{arXiv:1106.1238} [physics.ins-det]].



\bibitem{NOvA:2021nfi}
M.~A.~Acero \textit{et al.} [NOvA],
``{\it Improved measurement of neutrino oscillation parameters by the NOvA experiment},''
\href{https://doi.org/10.1103/PhysRevD.106.032004}{Phys. Rev. D \textbf{106}, no.3, 032004 (2022)}
[\href{https://arxiv.org/abs/2108.08219}{arXiv:2108.08219} [hep-ex]].

\bibitem{T2K:2023smv}
K.~Abe \textit{et al.} [T2K],
``{\it Measurements of neutrino oscillation parameters from the T2K experiment using $3.6\times 10^{21}$ protons on target},''
\href{https://doi.org/10.1140/epjc/s10052-023-11819-x}{Eur. Phys. J. C \textbf{83}, no.9, 782 (2023)}
[\href{https://arxiv.org/abs/2303.03222}{arXiv:2303.03222} [hep-ex]].




\bibitem{Abe:2011ts}
K.~Abe, T.~Abe, H.~Aihara, Y.~Fukuda, Y.~Hayato, K.~Huang, A.~K.~Ichikawa, M.~Ikeda, K.~Inoue and H.~Ishino, \textit{et al.}
``{\it Letter of Intent: The Hyper-Kamiokande Experiment --- Detector Design and Physics Potential ---},''
[\href{https://arxiv.org/abs/1109.3262}{arXiv:1109.3262} [hep-ex]].

\bibitem{DUNE:2015lol}
R.~Acciarri \textit{et al.} [DUNE],
``\textit{Long-Baseline Neutrino Facility (LBNF) and Deep Underground Neutrino Experiment (DUNE): Conceptual Design Report, Volume 2: The Physics Program for DUNE at LBNF},''
[\href{https://arxiv.org/abs/1512.06148}{arXiv:1512.06148} [physics.ins-det]].

\bibitem{Hyper-Kamiokande:2018ofw}
K.~Abe \textit{et al.} [Hyper-Kamiokande],
``\textit{Hyper-Kamiokande Design Report},''
[\href{https://arxiv.org/abs/1805.04163}{arXiv:1805.04163} [physics.ins-det]].

%\cite{DUNE:2021mtg}
\bibitem{DUNE:2021mtg}
A.~Abud Abed \textit{et al.} [DUNE],
``{\it Low exposure long-baseline neutrino oscillation sensitivity of the DUNE experiment},''
\href{https://doi.org/10.1103/PhysRevD.105.072006}{Phys. Rev. D \textbf{105}, no.7, 072006 (2022)}
[\href{https://arxiv.org/abs/2109.01304}{arXiv:2109.01304} [hep-ex]].

\bibitem{ESSnuSB:2021azq}
  A.~Alekou \textit{et al.} [ESSnuSB],
  ``{\it Updated physics performance of the ESSnuSB experiment: ESSnuSB collaboration},''
  \href{http://dx.doi.org/10.1140/epjc/s10052-021-09845-8}
  {Eur. Phys. J. C \textbf{81}, no.12, 1130 (2021)}
  [\href{http://arxiv.org/abs/2107.07585}{arXiv:2107.07585} [hep-ex]].

\bibitem{Hyper-Kamiokande:2016srs}
K.~Abe \textit{et al.} [Hyper-Kamiokande],
``{\it Physics potentials with the second Hyper-Kamiokande detector in Korea},''
\href{https://doi.org/10.1093/ptep/pty044}{
PTEP \textbf{2018}, no.6, 063C01 (2018)
}
[\href{https://arxiv.org/abs/1611.06118}{arXiv:1611.06118} [hep-ex]].

\bibitem{Tang:2019wsv}
J.~Tang, S.~Vihonen and T.~C.~Wang,
``{\it Precision measurements on $\delta_\text{CP}$ in MOMENT},''
\href{https://doi.org/10.1007/JHEP12(2019)130}{JHEP \textbf{12}, 130 (2019)}
[\href{https://arxiv.org/abs/1909.01548}{arXiv:1909.01548} [hep-ph]].

\bibitem{Akindinov:2019flp}
A.~V.~Akindinov, E.~G.~Anassontzis, G.~Anton, M.~Ardid, J.~Aublin, B.~Baret, V.~Bertin, S.~Bourret, C.~Bozza and M.~Bruchner, \textit{et al.}
``{\it Letter of Interest for a Neutrino Beam from Protvino to KM3NeT/ORCA},''
\href{https://doi.org/10.1140/epjc/s10052-019-7259-5}{Eur. Phys. J. C \textbf{79}, no.9, 758 (2019)}
[\href{https://arxiv.org/abs/1902.06083}{arXiv:1902.06083} [physics.ins-det]].

\bibitem{Alonso:2010fs}
J.~Alonso, F.~T.~Avignone, W.~A.~Barletta, R.~Barlow, H.~T.~Baumgartner, A.~Bernstein, E.~Blucher, L.~Bugel, L.~Calabretta and L.~Camilleri, \textit{et al.}
``{\it Expression of Interest for a Novel Search for CP Violation in the Neutrino Sector: DAEdALUS},''
[\href{https://arxiv.org/abs/1006.0260}{arXiv:1006.0260} [physics.ins-det]].

\bibitem{Evslin:2015pya}
Jarah Evslin, Shao-Feng Ge and Kaoru Hagiwara,
``{\it The leptonic CP phase from T2(H)K and \ensuremath{\mu}$^{+}$ decay at rest},''
\href{http://dx.doi.org/10.1007/JHEP02(2016)137}{JHEP \textbf{02}, 137 (2016)}
[\href{http://arxiv.org/abs/1506.05023}{arXiv:1506.05023} [hep-ph]];
%
  Shao-Feng Ge
  ``{\it Measuring the Leptonic Dirac CP Phase with TNT2K},''
  [\href{http://arxiv.org/abs/1704.08518}{arXiv:1704.08518} [hep-ph]]
  at \href{https://indico.ph.qmul.ac.uk/indico/conferenceDisplay.py?confId=112}
  {NuPhys2016: Prospects in Neutrino Physics},
  Barbican Centre, London, UK, December 12-14, 2016;
%


\bibitem{Ciuffoli:2014ika}
E.~Ciuffoli, J.~Evslin and X.~Zhang,
``{\it The Leptonic CP Phase from Muon Decay at Rest with Two Detectors},''
\href{https://doi.org/10.1007/JHEP12(2014)051}{
JHEP \textbf{12}, 051 (2014)
}
[\href{https://arxiv.org/abs/1401.3977}{
arXiv:1401.3977} [hep-ph]].

\bibitem{Ciuffoli:2015uta}
E.~Ciuffoli, J.~Evslin and F.~Zhao,
``{\it Neutrino Physics with Accelerator Driven Subcritical Reactors},''
\href{https://doi.org/10.1007/JHEP01(2016)004}{
JHEP \textbf{01}, 004 (2016)}
[\href{https://arxiv.org/abs/1509.03494}{arXiv:1509.03494} [hep-ph]].

\bibitem{Smirnov:2018ywm}
M.~V.~Smirnov, Z.~J.~Hu, S.~J.~Li and J.~J.~Ling,
``{\it The possibility of leptonic CP-violation measurement with JUNO},''
\href{https://doi.org/10.1016/j.nuclphysb.2018.05.003}{
Nucl. Phys. B \textbf{931}, 437-445 (2018)
}
[\href{https://arxiv.org/abs/1802.03677}{arXiv:1802.03677} [hep-ph]].

\bibitem{Ge:2022iac}
Shao-Feng Ge, Chui-Fan Kong and Pedro Pasquini,
``{\it Improving CP measurement with THEIA and muon decay at rest},''
\href{https://doi.org/10.1140/epjc/s10052-022-10478-8}{Eur. Phys. J. C \textbf{82}, no.6, 572 (2022)}
[\href{https://arxiv.org/abs/2202.05038}{arXiv:2202.05038} [hep-ph]].

\bibitem{Razzaque:2014vba}
S.~Razzaque and A.~Y.~Smirnov,
``{\it Super-PINGU for measurement of the leptonic CP-phase with atmospheric neutrinos},''
\href{https://doi.org/10.1007/JHEP05(2015)139}{
JHEP \textbf{05}, 139 (2015)}
[\href{https://arxiv.org/abs/1406.1407}{arXiv:1406.1407} [hep-ph]].

\bibitem{Razzaque:2015fea}
S.~Razzaque and A.~Y.~Smirnov,
``{\it Super-PINGU for measuring CP violation},''
\href{https://doi.org/10.1016/j.nuclphysbps.2015.06.046}{
Nucl. Part. Phys. Proc. \textbf{265-266}, 183-185 (2015)}
[\href{https://arxiv.org/abs/1501.03145}{arXiv:1501.03145} [hep-ph]].

\bibitem{Hofestadt:2019whx}
J.~Hofest\"adt, M.~Bruchner and T.~Eberl,
``{\it Super-ORCA: Measuring the leptonic CP-phase with Atmospheric Neutrinos and Beam Neutrinos},''
\href{https://doi.org/10.22323/1.358.0911}{
PoS \textbf{ICRC2019}, 911 (2020)
}
[\href{https://arxiv.org/abs/1907.12983}{arXiv:1907.12983} [hep-ex]].

\bibitem{JUNO:2015zny}
F.~An \textit{et al.} [JUNO],
``{\it Neutrino Physics with JUNO},''
\href{https://doi.org/10.1088/0954-3899/43/3/030401}{
J. Phys. G \textbf{43}, no.3, 030401 (2016)
}
[\href{https://arxiv.org/abs/1507.05613}{arXiv:1507.05613} [physics.ins-det]].

\bibitem{Kelly:2019itm}
K.~J.~Kelly, P.~A.~Machado, I.~Martinez Soler, S.~J.~Parke and Y.~F.~Perez Gonzalez,
``{\it Sub-GeV Atmospheric Neutrinos and CP-Violation in DUNE},''
\href{https://doi.org/10.1103/PhysRevLett.123.081801}{
Phys. Rev. Lett. \textbf{123}, no.8, 081801 (2019)}
[\href{https://arxiv.org/abs/1904.02751}{arXiv:1904.02751} [hep-ph]].

\bibitem{deSalas:2020pgw}
P.~F.~de Salas, D.~V.~Forero, S.~Gariazzo, P.~Mart\'\i{}nez-Mirav\'e, O.~Mena, C.~A.~Ternes, M.~T\'ortola and J.~W.~F.~Valle,
``{\it 2020 global reassessment of the neutrino oscillation picture},''
\href{http://dx.doi.org/10.1007/JHEP02(2021)071}{JHEP \textbf{02} (2021), 071}
[\href{http://arxiv.org/abs/2006.11237}{arXiv:2006.11237} [hep-ph]].
See tables in the Valencia neutrino \href{http://globalfit.astroparticles.es/}{ global fit}.

\bibitem{Gonzalez-Garcia:2021dve}
M.~C.~Gonzalez-Garcia, M.~Maltoni and T.~Schwetz,
``\textit{NuFIT: Three-Flavour Global Analyses of Neutrino Oscillation Experiments},''
\href{https://doi.org/10.3390/universe7120459}{
Universe \textbf{7} (2021) no.12, 459
}
[\href{https://arxiv.org/abs/2111.03086}{arXiv:2111.03086} [hep-ph]].

\bibitem{DeRomeri:2016qwo}
V.~De Romeri, E.~Fernandez-Martinez and M.~Sorel,
``\textit{Neutrino oscillations at DUNE with improved energy reconstruction},''
\href{https://doi.org/10.1007/JHEP09(2016)030}
{JHEP \textbf{09} (2016), 030}
[\href{https://arxiv.org/abs/1607.00293 }{arXiv:1607.00293 } [hep-ph]].

\bibitem{Raut:2017dbh}
S.~K.~Raut,
``\textit{Matter effects at the T2HK and T2HKK experiments},''
\href{https://doi.org/10.1103/PhysRevD.96.075029}
{Phys. Rev. D \textbf{96} (2017) no.7, 075029}
[\href{https://arxiv.org/abs/1703.07136}{arXiv:1703.07136} [hep-ph]].

\bibitem{Kelly:2018kmb}
K.~J.~Kelly and S.~J.~Parke,
``\textit{Matter Density Profile Shape Effects at DUNE},''
\href{https://doi.org/10.1103/PhysRevD.98.015025}
{Phys. Rev. D \textbf{98} (2018) no.1, 015025}
[\href{https://arxiv.org/abs/1802.06784}{arXiv:1802.06784} [hep-ph]].

\bibitem{King:2020ydu}
S.~F.~King, S.~Molina Sedgwick, S.~J.~Parke and N.~W.~Prouse,
``\textit{Effects of matter density profiles on neutrino oscillations for T2HK and T2HKK},''
\href{https://doi.org/10.1103/PhysRevD.101.076019}
{Phys. Rev. D \textbf{101} (2020), 076019}
[\href{https://arxiv.org/abs/2001.05505}{arXiv:2001.05505} [hep-ph]].

\bibitem{Farzan:2017xzy}
Y.~Farzan and M.~Tortola,
``{\it Neutrino oscillations and Non-Standard Interactions},''
\href{http://dx.doi.org/10.3389/fphy.2018.00010}
{Front. in Phys. \textbf{6}, 10 (2018)}
[\href{http://arxiv.org/abs/1710.09360}{arXiv:1710.09360} [hep-ph]].

\bibitem{Proceedings:2019qno}
P.~S.~Bhupal Dev, K.~S.~Babu, P.~B.~Denton, P.~A.~N.~Machado, C.~A.~Arg\"uelles, J.~L.~Barrow, S.~S.~Chatterjee, M.~C.~Chen, A.~de Gouv\^ea and B.~Dutta, \textit{et al.}
``{\it Neutrino Non-Standard Interactions: A Status Report},''
\href{https://doi.org/10.21468/SciPostPhysProc.2.001}{SciPost Phys. Proc. \textbf{2}, 001 (2019)}
[\href{https://arxiv.org/abs/1907.00991}{arXiv:1907.00991} [hep-ph]].

\bibitem{Ge:2016dlx}
  Shao-Feng Ge and Alexei Y. Smirnov,
  ``{\it Non-standard interactions and the CP phase measurements in neutrino oscillations at low energies},''
  \href{http://dx.doi.org/10.1007/JHEP10(2016)138}
  {JHEP \textbf{10}, 138 (2016)}
  [\href{http://arxiv.org/abs/1607.08513}{arXiv:1607.08513} [hep-ph]].

\bibitem{GeProceedings}
  Shao-Feng Ge,
  ``{\it The Leptonic CP Measurement and New Physics Alternatives},''
  \href{https://doi.org/10.22323/1.369.0108}
       {PoS NuFact2019 (2020) 108}
  at
  \href{https://indico.cern.ch/event/773605/}
       {The 21st International Workshop on Neutrinos from Accelerators (NUFACT2019)},
       Daegu, Korea, August 29, 2019;
%
  ``{\it New Physics with Scalar and Dark Non-Standard Interactions in Neutrino Oscillation,}''
  \href{https://doi.org/10.1088/1742-6596/1468/1/012125}
       {J.Phys.Conf.Ser. 1468 (2020) 1, 012125}
  at
  \href{http://taup2019.icrr.u-tokyo.ac.jp/}
       {The 16th International Conference on Topics in Astroparticle and Underground Physics},
      Toyama, Japan, September 10, 2019.

\bibitem{Bakhti:2020fde}
P.~Bakhti and M.~Rajaee,
``{\it Sensitivities of future reactor and long-baseline neutrino experiments to NSI},''
\href{https://doi.org/10.1103/PhysRevD.103.075003}{Phys. Rev. D \textbf{103}, no.7, 075003 (2021)}
[\href{https://arxiv.org/abs/2010.12849}{arXiv:2010.12849} [hep-ph]].

\bibitem{Chatterjee:2021wac}
S.~S.~Chatterjee, P.~S.~B.~Dev and P.~A.~N.~Machado,
``{\it Impact of improved energy resolution on DUNE sensitivity to neutrino non-standard interactions},''
\href{https://doi.org/10.1007/JHEP08(2021)163}{JHEP \textbf{08}, 163 (2021)}
[\href{https://arxiv.org/abs/2106.04597}{arXiv:2106.04597} [hep-ph]].

\bibitem{Medhi:2021wxj}
A.~Medhi, D.~Dutta and M.~M.~Devi,
``{\it Exploring the effects of scalar non standard interactions on the CP violation sensitivity at DUNE},''
\href{https://doi.org/10.1007/JHEP06(2022)129}{JHEP \textbf{06}, 129 (2022)}
[\href{https://arxiv.org/abs/2111.12943}{arXiv:2111.12943} [hep-ph]].

\bibitem{Medhi:2022qmu}
A.~Medhi, M.~M.~Devi and D.~Dutta,
``{\it Imprints of scalar NSI on the CP-violation sensitivity using synergy among DUNE, T2HK and T2HKK},''
\href{https://doi.org/10.1007/JHEP01(2023)079}{JHEP \textbf{01}, 079 (2023)}
[\href{https://arxiv.org/abs/2209.05287}{arXiv:2209.05287} [hep-ph]].

\bibitem{Diaz:2014yva}
J.~S.~Diaz,
``{\it Neutrinos as probes of Lorentz invariance},''
\href{https://doi.org/10.1155/2014/962410}{Adv. High Energy Phys. \textbf{2014}, 962410 (2014)}
[\href{https://arxiv.org/abs/1406.6838}{arXiv:1406.6838} [hep-ph]].

\bibitem{Torri:2020dec}
M.~D.~C.~Torri,
``{\it Neutrino Oscillations and Lorentz Invariance Violation},''
\href{https://doi.org/10.3390/universe6030037}{Universe \textbf{6}, no.3, 37 (2020)}
[\href{https://arxiv.org/abs/2110.09186}{arXiv:2110.09186} [hep-ph]].

\bibitem{Lin:2021cst}
H.~X.~Lin, J.~Tang, S.~Vihonen and P.~Pasquini,
``\textit{Nonminimal Lorentz invariance violation in light of the muon anomalous magnetic moment and long-baseline neutrino oscillation data},''
\href{https://doi.org/10.1103/PhysRevD.105.096029}{Phys. Rev. D \textbf{105} (2022) no.9, 096029}
[\href{https://arxiv.org/abs/2111.14336}{arXiv:2111.14336} [hep-ph]].

\bibitem{Majhi:2019tfi}
R.~Majhi, S.~Chembra and R.~Mohanta,
``{\it Exploring the effect of Lorentz invariance violation with the currently running long-baseline experiments},''
\href{https://doi.org/10.1140/epjc/s10052-020-7963-1}{Eur. Phys. J. C \textbf{80}, no.5, 364 (2020)}
[\href{https://arxiv.org/abs/1907.09145}{arXiv:1907.09145} [hep-ph]].

\bibitem{KumarAgarwalla:2019gdj}
S.~Kumar Agarwalla and M.~Masud,
``{\it Can Lorentz invariance violation affect the sensitivity of deep underground neutrino experiment?},''
\href{https://doi.org/10.1140/epjc/s10052-020-8303-1}{Eur. Phys. J. C \textbf{80}, no.8, 716 (2020)}
[\href{https://arxiv.org/abs/1912.13306}{arXiv:1912.13306} [hep-ph]].

\bibitem{Fiza:2022xfw}
N.~Fiza, N.~R.~Khan Chowdhury and M.~Masud,
``{\it Investigating Lorentz Invariance Violation with the long baseline experiment P2O},''
\href{https://doi.org/10.1007/JHEP01(2023)076}{JHEP \textbf{01}, 076 (2023)}
[\href{https://arxiv.org/abs/2206.14018}{arXiv:2206.14018} [hep-ph]].

\bibitem{Forero:2011pc}
D.~V.~Forero, S.~Morisi, M.~Tortola and J.~W.~F.~Valle,
``{\it Lepton flavor violation and non-unitary lepton mixing in low-scale type-I seesaw},''
\href{https://doi.org/10.1007/JHEP09(2011)142}{JHEP \textbf{09}, 142 (2011)}
[\href{https://arxiv.org/abs/1107.6009}{arXiv:1107.6009} [hep-ph]].

\bibitem{Escrihuela:2015wra}
F.~J.~Escrihuela, D.~V.~Forero, O.~G.~Miranda, M.~Tortola and J.~W.~F.~Valle,
``{\it On the description of nonunitary neutrino mixing},''
\href{https://doi.org/10.1103/PhysRevD.92.053009}{Phys. Rev. D \textbf{92}, no.5, 053009 (2015)}
[erratum: \href{https://doi.org/10.1103/PhysRevD.93.119905}{Phys. Rev. D \textbf{93}, no.11, 119905 (2016)}]
[\href{https://arxiv.org/abs/1503.08879}{arXiv:1503.08879} [hep-ph]].

\bibitem{Ge:2016xya}
Shao-Feng Ge, Pedro Pasquini, M.~Tortola and J.~W.~F.~Valle,
``{\it Measuring the leptonic CP phase in neutrino oscillations with nonunitary mixing},''
\href{https://doi.org/10.1103/PhysRevD.95.033005}{Phys. Rev. D \textbf{95}, no.3, 033005 (2017)}
[\href{https://arxiv.org/abs/1605.01670}{arXiv:1605.01670} [hep-ph]].

\bibitem{Pasquini:2017zwk}
  P.~Pasquini, S.~F.~Ge, M.~A.~T\'ortola and J.~W.~F.~Valle,
  ``{\it Measuring the Leptonic CP Phase in Neutrino Oscillations with Non-Unitary Mixing},''
  \href{https://doi.org/10.22323/1.283.0026}
   {PoS \textbf{NOW2016}, 026 (2017)}
  at the
   \href{https://web2.ba.infn.it/now/now2016/}
        {Neutrino Oscillation Workshop (NOW) 2016},
       Otranto, Lecce, Italy, September 4-11, 2016.

\bibitem{Escrihuela:2016ube}
F.~J.~Escrihuela, D.~V.~Forero, O.~G.~Miranda, M.~T\'ortola and J.~W.~F.~Valle,
``{\it Probing CP violation with non-unitary mixing in long-baseline neutrino oscillation experiments: DUNE as a case study},''
\href{https://doi.org/10.1088/1367-2630/aa79ec}{New J. Phys. \textbf{19}, no.9, 093005 (2017)}
[\href{https://arxiv.org/abs/1612.07377}{arXiv:1612.07377} [hep-ph]].

\bibitem{Forero:2021azc}
D.~V.~Forero, C.~Giunti, C.~A.~Ternes and M.~Tortola,
``{\it Nonunitary neutrino mixing in short and long-baseline experiments},''
\href{https://doi.org/10.1103/PhysRevD.104.075030}{Phys. Rev. D \textbf{104}, no.7, 075030 (2021)}
[\href{https://arxiv.org/abs/2103.01998}{arXiv:2103.01998} [hep-ph]].

\bibitem{Martinez-Soler:2018lcy}
I.~Martinez-Soler and H.~Minakata,
``{\it Standard versus Non-Standard CP Phases in Neutrino Oscillation in Matter with Non-Unitarity},''
\href{https://doi.org/10.1093/ptep/ptaa062}{
PTEP \textbf{2020}, no.6, 063B01 (2020)
}
[\href{https://arxiv.org/abs/1806.10152}{arXiv:1806.10152} [hep-ph]].

\bibitem{Huang:2023nqf}
J.~Huang and S.~Zhou,
``{\it The Mikheyev-Smirnov-Wolfenstein Matter Potential at the One-loop Level in the Standard Model},''
[\href{https://arxiv.org/abs/2307.04685}{arXiv:2307.04685} [hep-ph]].

\bibitem{Casas:1999tg}
J.~A.~Casas, J.~R.~Espinosa, A.~Ibarra and I.~Navarro,
``{\it General RG equations for physical neutrino parameters and their phenomenological implications},''
\href{https://doi.org/10.1016/S0550-3213(99)00781-6}{Nucl. Phys. B \textbf{573}, 652-684 (2000)}
[\href{https://arxiv.org/abs/hep-ph/9910420}{arXiv:hep-ph/9910420} [hep-ph]].

\bibitem{Chankowski:2001mx}
P.~H.~Chankowski and S.~Pokorski,
``{\it Quantum corrections to neutrino masses and mixing angles},''
\href{https://doi.org/10.1142/S0217751X02006109}{Int. J. Mod. Phys. A \textbf{17}, 575-614 (2002)}
[\href{https://arxiv.org/abs/hep-ph/0110249}{arXiv:hep-ph/0110249} [hep-ph]].

\bibitem{Antusch:2003kp}
S.~Antusch, J.~Kersten, M.~Lindner and M.~Ratz,
``{\it Running neutrino masses, mixings and CP phases: Analytical results and phenomenological consequences},''
\href{https://doi.org/10.1016/j.nuclphysb.2003.09.050}{Nucl. Phys. B \textbf{674}, 401-433 (2003)}
[\href{https://arxiv.org/abs/hep-ph/0305273}{arXiv:hep-ph/0305273} [hep-ph]].

\bibitem{Ray:2010rz}
S.~Ray,
``\textit{Renormalization group evolution of neutrino masses and mixing in seesaw models: A Review},''
\href{https://doi.org/0.1142/S0217751X10049839}{Int. J. Mod. Phys. A \textbf{25}, 4339-4384 (2010)}
[\href{https://arxiv.org/abs/1005.1938}{arXiv:1005.1938} [hep-ph]].

\bibitem{Ohlsson:2013xva}
T.~Ohlsson and S.~Zhou,
``\textit{Renormalization group running of neutrino parameters},''
\href{https://doi.org/10.1038/ncomms6153}{Nature Commun. \textbf{5} (2014), 5153}
[\href{https://arxiv.org/abs/1311.3846}{arXiv:1311.3846} [hep-ph]].

\bibitem{Babu:2021cxe}
K.~S.~Babu, V.~Brdar, A.~de Gouv\^ea and P.~A.~N.~Machado,
``\textit{Energy-dependent neutrino mixing parameters at oscillation experiments},''
\href{https://doi.org/10.1103/PhysRevD.105.115014}{Phys. Rev. D \textbf{105}, no.11, 115014 (2022)}
[\href{https://arxiv.org/abs/2108.11961}{arXiv:2108.11961} [hep-ph]].

\bibitem{Gell-Mann:1954yli}
M.~Gell-Mann and F.~E.~Low,
``\textit{Quantum electrodynamics at small distances},''
\href{https://doi.org/10.1103/PhysRev.95.1300}{Phys. Rev. \textbf{95}, 1300-1312 (1954)}

\bibitem{Wu:2013ei}
X.~G.~Wu, S.~J.~Brodsky and M.~Mojaza,
``\textit{The Renormalization Scale-Setting Problem in QCD},''
\href{https://doi.org/10.1016/j.ppnp.2013.06.001}{Prog. Part. Nucl. Phys. \textbf{72}, 44-98 (2013)}
[\href{https://arxiv.org/abs/1302.0599}{arXiv:1302.0599} [hep-ph]].


\bibitem{Bustamante:2010bf}
M.~Bustamante, A.~M.~Gago and J.~Jones Perez,
``{\it SUSY Renormalization Group Effects in Ultra High Energy Neutrinos},''
\href{https://doi.org/10.1007/JHEP05(2011)133}{JHEP \textbf{05}, 133 (2011)}
[\href{https://arxiv.org/abs/1012.2728}{arXiv:1012.2728} [hep-ph]].

\bibitem{Babu:2022non}
K.~S.~Babu, V.~Brdar, A.~de Gouv\^ea and P.~A.~N.~Machado,
``\textit{Addressing the short-baseline neutrino anomalies with energy-dependent mixing parameters},''
\href{https://doi.org/10.1103/PhysRevD.107.015017}{Phys. Rev. D \textbf{107}, no.1, 015017 (2023)}
[\href{https://arxiv.org/abs/2209.00031}{arXiv:2209.00031} [hep-ph]].

\bibitem{Babu:1993qv}
K.~S.~Babu, C.~N.~Leung and J.~T.~Pantaleone,
``\textit{Renormalization of the neutrino mass operator},''
\href{https://doi.org/10.1016/0370-2693(93)90801-N}{Phys. Lett. B \textbf{319}, 191-198 (1993)}
[\href{https://arxiv.org/abs/hep-ph/9309223}{arXiv:hep-ph/9309223} [hep-ph]].



\bibitem{Chankowski:1993tx}
P.~H.~Chankowski and Z.~Pluciennik,
``\textit{Renormalization group equations for seesaw neutrino masses},''
\href{https://doi.org/10.1016/0370-2693(93)90330-K}{Phys. Lett. B \textbf{316} (1993), 312-317}
[\href{https://arxiv.org/abs/hep-ph/9306333}{arXiv:hep-ph/9306333} [hep-ph]].

\bibitem{Grimus:1996av}
W.~Grimus and P.~Stockinger,
``{\it Real oscillations of virtual neutrinos},''
\href{https://doi.org/10.1103/PhysRevD.54.3414}{Phys. Rev. D \textbf{54}, 3414-3419 (1996)}
[\href{https://arxiv.org/abs/hep-ph/9603430}{arXiv:hep-ph/9603430} [hep-ph]].


\bibitem{Falkowski:2019kfn}
A.~Falkowski, M.~Gonz\'alez-Alonso and Z.~Tabrizi,
``{\it Consistent QFT description of non-standard neutrino interactions},''
\href{https://doi.org/10.1007/JHEP11(2020)048}{JHEP \textbf{11} (2020), 048}
[\href{https://arxiv.org/abs/1910.02971}{arXiv:1910.02971} [hep-ph]].

\bibitem{Giunti:1993se}
C.~Giunti, C.~W.~Kim, J.~A.~Lee and U.~W.~Lee,
``{\it On the treatment of neutrino oscillations without resort to weak eigenstates},''
\href{https://doi.org/10.1103/PhysRevD.48.4310}{Phys. Rev. D \textbf{48}, 4310-4317 (1993)}
[\href{https://arxiv.org/abs/hep-ph/9305276}{arXiv:hep-ph/9305276} [hep-ph]].

\bibitem{Beuthe:2001rc}
M.~Beuthe,
``\textit{Oscillations of neutrinos and mesons in quantum field theory},''
\href{https://doi.org/10.1016/S0370-1573(02)00538-0}{Phys. Rept. \textbf{375} (2003), 105-218}
[\href{https://arxiv.org/abs/hep-ph/0109119}{arXiv:hep-ph/0109119} [hep-ph]].

\bibitem{DUNE:2020ypp}
B.~Abi \textit{et al.} [DUNE],
``{\it Deep Underground Neutrino Experiment (DUNE), Far Detector Technical Design Report, Volume II: DUNE Physics},''
[\href{https://arxiv.org/abs/2002.03005}{arXiv:2002.03005} [hep-ex]].

\bibitem{Andreopoulos:2009rq}
C.~Andreopoulos, A.~Bell, D.~Bhattacharya, F.~Cavanna, J.~Dobson, S.~Dytman, H.~Gallagher, P.~Guzowski, R.~Hatcher and P.~Kehayias, \textit{et al.}
``\textit{The GENIE Neutrino Monte Carlo Generator},''
\href{https://doi.org/10.1016/j.nima.2009.12.009}{Nucl. Instrum. Meth. A \textbf{614}, 87-104 (2010)}
[\href{https://arxiv.org/abs/0905.2517}{arXiv:0905.2517} [hep-ph]].

\bibitem{Andreopoulos:2015wxa}
C.~Andreopoulos, C.~Barry, S.~Dytman, H.~Gallagher, T.~Golan, R.~Hatcher, G.~Perdue and J.~Yarba,
``\textit{The GENIE Neutrino Monte Carlo Generator: Physics and User Manual},''
[\href{https://arxiv.org/abs/1510.05494}{arXiv:1510.05494} [hep-ph]].


\bibitem{Friedland:2018vry}
A.~Friedland and S.~W.~Li,
``\textit{Understanding the energy resolution of liquid argon neutrino detectors},''
\href{https://doi.org/10.1103/PhysRevD.99.036009}{Phys. Rev. D \textbf{99}, no.3, 036009 (2019)}
[\href{https://arxiv.org/abs/1811.06159}{arXiv:1811.06159} [hep-ph]].



\bibitem{Ge:2013ffa}
Shao-Feng Ge and Kaoru Hagiwara,
``\textit{Physics Reach of Atmospheric Neutrino Measurements at PINGU},''
\href{https://doi.org/10.1007/JHEP09(2014)024}{JHEP \textbf{09}, 024 (2014)}
[\href{https://arxiv.org/abs/1312.0457}{arXiv:1312.0457} [hep-ph]].

\bibitem{Huber:2004ka}
P.~Huber, M.~Lindner and W.~Winter,
``{\it Simulation of long-baseline neutrino oscillation experiments with GLoBES (General Long Baseline Experiment Simulator)},''
\href{http://dx.doi.org/10.1016/j.cpc.2005.01.003}{Comput. Phys. Commun. \textbf{167}, 195 (2005)}
[\href{http://arxiv.org/abs/hep-ph/0407333}{arXiv:hep-ph/0407333} [hep-ph]].

\bibitem{Huber:2007ji}
P.~Huber, J.~Kopp, M.~Lindner, M.~Rolinec and W.~Winter,
``{\it New features in the simulation of neutrino oscillation experiments with GLoBES 3.0: General Long Baseline Experiment Simulator},''
\href{http://dx.doi.org/10.1016/j.cpc.2007.05.004}{Comput. Phys. Commun. \textbf{177}, 432-438 (2007)}
[\href{http://arxiv.org/abs/hep-ph/0701187}{arXiv:hep-ph/0701187} [hep-ph]].

\bibitem{Winter:2013ema}
W.~Winter,
``\textit{Neutrino mass hierarchy determination with IceCube-PINGU},''
\href{https://doi.org/10.1103/PhysRevD.88.013013}{Phys. Rev. D \textbf{88}, no.1, 013013 (2013)}
[\href{https://arxiv.org/abs/1305.5539}{arXiv:1305.5539} [hep-ph]].


\bibitem{Winter:2015zwx}
W.~Winter,
``\textit{Atmospheric Neutrino Oscillations for Earth Tomography},''
\href{https://doi.org/10.1016/j.nuclphysb.2016.03.033}{Nucl. Phys. B \textbf{908}, 250-267 (2016)}
[\href{https://arxiv.org/abs/1511.05154}{arXiv:1511.05154} [hep-ph]].

\bibitem{Abi:2021arg}
B.~Abi \textit{et al.} [DUNE],
``{\it Experiment Simulation Configurations Approximating DUNE TDR},''
[\href{http://arxiv.org/abs/2103.04797}{arXiv:2103.04797} [hep-ex]].

\bibitem{ICARUS:2013cwr}
M.~Antonello \textit{et al.} [ICARUS],
``{\it Search for anomalies in the ${\nu}_e$ appearance from a ${\nu}_{\mu}$ beam},''
\href{https://doi.org/10.1140/epjc/s10052-013-2599-z}{Eur. Phys. J. C \textbf{73}, 2599 (2013)}
[\href{https://arxiv.org/abs/1307.4699}{arXiv:1307.4699} [hep-ex]].


\bibitem{CHARMII:1994rnc}
P.~Vilain \textit{et al.} [CHARM II],
``{\it Search for muon to electron-neutrino oscillations},''
\href{https://doi.org/10.1007/BF01957768}{Z. Phys. C \textbf{64}, 539-544 (1994)}

\bibitem{NOMAD:2003mqg}
P.~Astier \textit{et al.} [NOMAD],
``\textit{Search for nu(mu) ---\ensuremath{>} nu(e) oscillations in the NOMAD experiment},''
\href{https://doi.org/10.1016/j.physletb.2003.07.029}{Phys. Lett. B \textbf{570}, 19-31 (2003)}
[\href{https://arxiv.org/abs/hep-ex/0306037}{arXiv:hep-ex/0306037} [hep-ex]].

\bibitem{NuTeV:2002daf}
S.~Avvakumov \textit{et al.} [NuTeV],
``\textit{A Search for $\nu_{\mu} \to \nu_e$ and $\bar{\nu}_{\mu} \to \bar \nu_e$ Oscillations at NuTeV},''
\href{https://doi.org/10.1103/PhysRevLett.89.011804}{Phys. Rev. Lett. \textbf{89}, 011804 (2002)}
[\href{https://arxiv.org/abs/hep-ex/0203018}{arXiv:hep-ex/0203018} [hep-ex]].

\bibitem{CHARM-II:1993nip}
M.~Gruwe \textit{et al.} [CHARM-II],
``{\it Search for muon-neutrino ---\ensuremath{>} tau-neutrino oscillation},''
\href{https://doi.org/10.1016/0370-2693(93)90962-H}{Phys. Lett. B \textbf{309}, 463-468 (1993)}

\bibitem{NOMAD:2001xxt}
P.~Astier \textit{et al.} [NOMAD],
``\textit{Final NOMAD results on muon-neutrino ---\ensuremath{>} tau-neutrino and electron-neutrino ---\ensuremath{>} tau-neutrino oscillations including a new search for tau-neutrino appearance using hadronic tau decays},''
\href{https://doi.org/10.1016/S0550-3213(01)00339-X}{Nucl. Phys. B \textbf{611}, 3-39 (2001)}
[\href{https://arxiv.org/abs/hep-ex/0106102}{arXiv:hep-ex/0106102} [hep-ex]].

\bibitem{CCFRNuTeV:1998gjj}
D.~Naples \textit{et al.} [CCFR/NuTeV],
``\textit{A High statistics search for neutrino(e) (anti-neutrino(e)) ---\ensuremath{>} neutrino(tau) (anti-neutrino(tau)) oscillations},''
\href{https://doi.org/10.1103/PhysRevD.59.031101}{Phys. Rev. D \textbf{59}, 031101 (1999)}
[\href{https://arxiv.org/abs/hep-ex/9809023}{arXiv:hep-ex/9809023} [hep-ex]].


\bibitem{NOMAD:2003owt}
P.~Astier \textit{et al.} [NOMAD],
``{\it Prediction of neutrino fluxes in the NOMAD experiment},''
\href{https://doi.org/10.1016/j.nima.2003.07.054}{Nucl. Instrum. Meth. A \textbf{515}, 800-828 (2003)}
[\href{https://arxiv.org/abs/hep-ex/0306022}{arXiv:hep-ex/0306022 [hep-ex]}].

\end{thebibliography}

\end{document}